\documentclass[useAMS,usenatbib]{mn2e}
  \usepackage{amsmath}
  \usepackage{graphicx}

  \title[DM fraction of low-mass lens cluster members]{Dark matter fraction of low-mass cluster members probed by galaxy-scale strong lensing}
  \author[W. G. Parry et al.]
    {W.~G.~Parry$^{1}$\thanks{E-mail: gruffyddparry@gmail.com, grillo@dark-cosmology.dk},
    C.~Grillo$^{1}\footnotemark[1]$,
    A.~Mercurio$^{2}$,
    I.~Balestra$^{3,4}$,
    P.~Rosati$^{5}$,
    L.~Christensen$^{1}$,
    \newauthor
    M.~Lombardi$^{6}$,
    G.~B.~Caminha$^{5}$,
    M.~Nonino$^{4}$,
    A.~M.~Koekemoer$^{7}$, and
    K.~Umetsu$^{8}$\\
    $^{1}${Dark Cosmology Centre, Niels Bohr Institute, University of Copenhagen, Juliane Maries Vej 30, DK-2100 Copenhagen, Denmark}\\
    $^{2}${INAF - Osservatorio Astronomico di Capodimonte, Via Moiariello 16, I-80131 Napoli, Italy}\\
    $^{3}${University Observatory Munich, Scheinerstrasse 1, D-81679 Munich, Germany}\\
    $^{4}${INAF - Osservatorio Astronomico di Trieste, via G. B. Tiepolo 11, I-34143, Trieste, Italy}\\
    $^{5}${Dipartimento di Fisica e Scienze della Terra, Universit\`a degli Studi di Ferrara, Via Saragat 1, I-44122 Ferrara, Italy}\\
    $^{6}${Dipartimento di Fisica, Universit\`a  degli Studi di Milano, via Celoria 16, I-20133 Milano, Italy}\\
    $^{7}${Space Telescope Science Institute, 3700 San Martin Drive, Baltimore, MD 21208, USA}\\
    $^{8}${Institute of Astronomy and Astrophysics, Academia Sinica, P.O. Box 23-141, Taipei 10617, Taiwan}}
  
\begin{document}

  \date{}

  \pagerange{\pageref{firstpage}--\pageref{lastpage}} \pubyear{2002}

  \maketitle

  \label{firstpage}

\begin{abstract}

  We present a strong lensing system, composed of 4 multiple images of a source at $z=2.387$, created by two lens galaxies, G1 and G2, belonging to the galaxy cluster MACS J1115.9$+$0129 at $z=0.353$. We use observations taken as part of the Cluster Lensing and Supernova survey with Hubble, and its spectroscopic follow-up programme at the Very Large Telescope, to estimate the total mass distributions of the two galaxies and the cluster through strong gravitational lensing models. We find that the total projected mass values within the half-light radii, $R_{e}$, of the two lens galaxies are $M_{\text{T,G1}}(<R_{e,\text{G1}})=(3.6\pm0.4)\times10^{10}M_{\odot}$ and $M_{\text{T,G2}}(<R_{e,\text{G2}})=(4.2\pm1.6)\times10^{10}M_{\odot}$. The effective velocity dispersion values of G1 and G2 are $(122\pm7)$~km~s$^{-1}$ and $(137\pm27)$~km~s$^{-1}$, respectively. We remark that these values are relatively low when compared to those of $\approx200-300$~km~s$^{-1}$, typical of lens galaxies found in the field by previous surveys. By fitting the spectral energy distributions of G1 and G2, we measure projected luminous over total mass fractions within $R_{e}$ of $0.11\pm0.03$, for G1, and $0.73\pm0.32$, for G2. The fact that the less massive galaxy, G1, is dark-matter dominated in its inner regions raises the question of whether the dark matter fraction in the core of early-type galaxies depends on their mass. Further investigating strong lensing systems will help us understand the influence that dark matter has on the structure and evolution of the inner regions of galaxies.

  \end{abstract}


\section{Introduction}

  Over the past 40 years, gravitational lensing has become a valuable astrophysical tool for detailed studies of the internal structure of galaxies (e.g., \citealt{Kochanek00,Treu10a,Barnabe11}) and galaxy clusters (e.g., \citealt{Broadhurst95,Zitrin11,Umetsu15}) and estimates of the values of the cosmological parameters (e.g., \citealt{Grillo08a,Schwab10,Suyu13}). Weak lensing has allowed us to measure the total mass profile in the external regions of galaxies (e.g., \citealt{Gavazzi07,Brimioulle13}) and galaxy clusters (e.g., \citealt{Umetsu14,Applegate14}), while strong lensing has given us some of the most accurate measurements of the total mass of galaxies (e.g., \citealt{Koopmans06,Grillo08b}) and galaxy clusters within their Einstein radii, $R_{\text{Ein}}$ (e.g., \citealt{Zitrin09,Richard10,Grillo15}).

  Since the lensing cross-section depends on the mass of a lens, so far the majority of strong lensing galaxies that have been investigated are massive. The lensing cross-section of galaxies is naturally enhanced in overdense environments, because of the mass contribution of the hosting group or cluster. Therefore, it is there that strong lensing systems around low-mass galaxies are more likely to be observed. These systems are usually complex and require careful analyses to properly take into account the different mass components. Nonetheless, by combining strong lensing with photometric and dynamical models, we can extend to these lenses our knowledge about their central amount of dark matter, DM.

  Recent surveys, like the Sloan Lens ACS survey, SLACS \citep{Bolton06}, and the Cluster Lensing And Supernova survey with Hubble, CLASH (PI: Postman; \citealt{Postman12}), have significantly increased the number of strong lenses observed within galaxy clusters and groups and have shown a number of low-mass lens galaxies with resolved multiple images. This paper takes advantage of the CLASH data collected using the \emph{Hubble Space Telescope} (\emph{HST}) Advanced Camera for Surveys (ACS) to study a lensing system in which the lens is a member galaxy of a cluster in the sample.

  CLASH is a programme which observed, between November 2010 and July 2013, 25 galaxy clusters in 16 passbands, from the near-UV to the near-IR, totalling 524 orbits of time on \emph{HST}. The spectroscopic follow-up programme, CLASH-VLT (186.A-0798, PI: Rosati; \citealt{Rosati14}), started in October 2010 and obtained data using the VIsible MultiObject Spectrograph (VIMOS) instrument at the Very Large Telescope (VLT) of the European Southern Observatory. The main goal of these programmes is to measure accurately the total mass profiles of a statistically significant sample of galaxy clusters through different total mass diagnostics.

  The use of strong gravitational lensing to study low-mass galaxies is relatively new (e.g., \citealt{Grillo14}, hereafter G14, and \citealt{S15}, hereafter S15) and it offers us the opportunity to test the interplay between ordinary, baryonic matter and DM at different mass scales.

  The aim of this work is to extend strong lensing analyses to low-mass galaxies. This paper is structured as follows. Section~\ref{sec:data} introduces the lensing system and the available data. Section~\ref{sec:modelling} describes our photometric and lensing models. In Section~\ref{sec:discussion} we discuss the results on the galaxy DM fractions, and compare them with those of previous strong lensing studies. The standard $\Lambda$CDM model is adopted throughout this paper, where $\Omega_m=0.3$, $\Omega_{\Lambda}=0.7$ and $H_0=70\text{ km s}^{-1}\text{Mpc}^{-1}$. At the cluster redshift, $1''$ corresponds to 4.97 kpc.

\section{The lensing system}\label{sec:data}

\begin{figure*}
    \centering
    \includegraphics[width=0.8\textwidth]{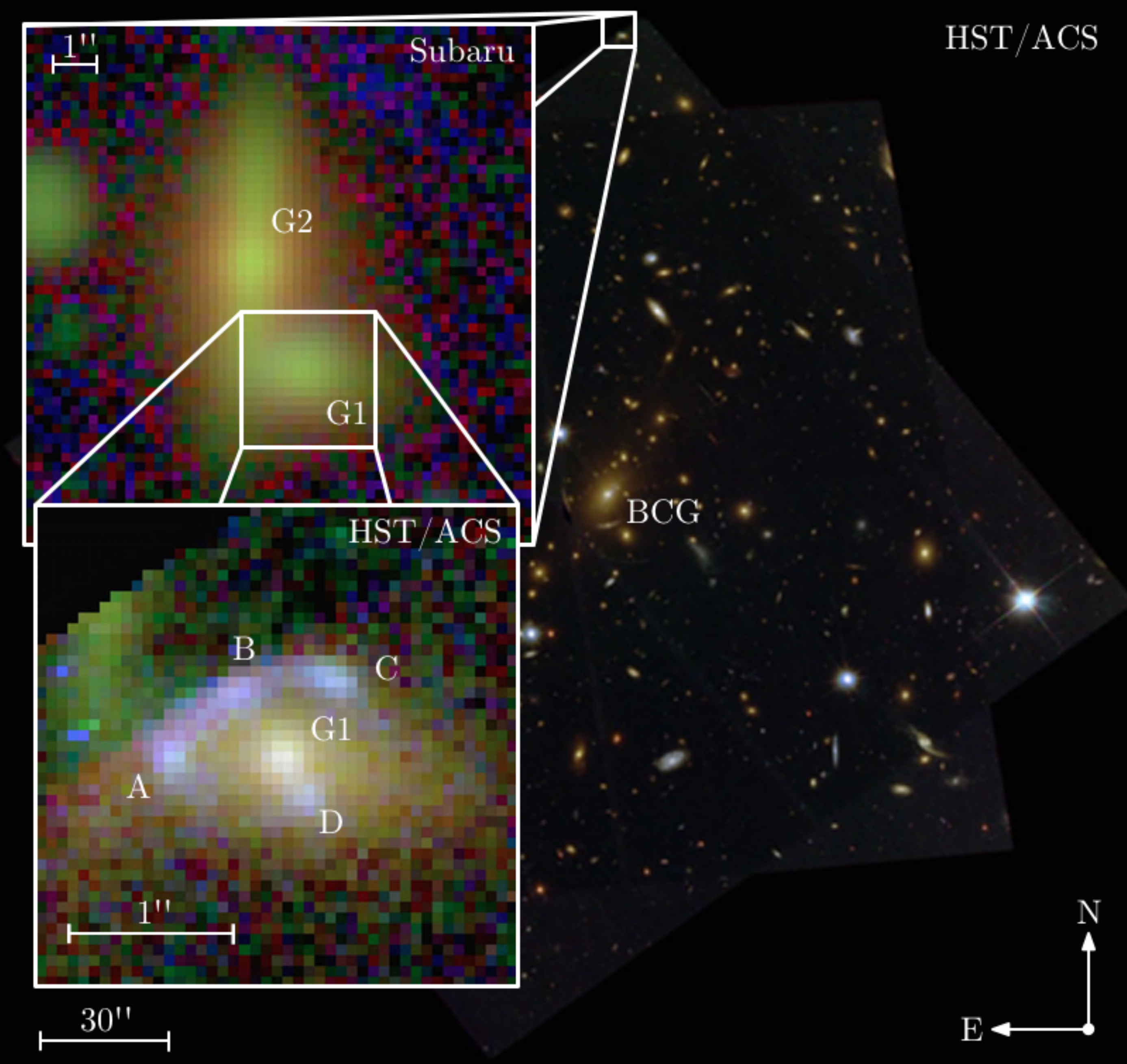}
    \caption{Colour-composite images of the strong lensing system obtained with Subaru and \emph{HST/ACS}. The high angular resolution of the \emph{HST/ACS} data allows us to resolve the multiple images (A, B, C and D) of the background source. The two main lenses, G1 and G2, are members of the galaxy cluster MACS~J1115.9+0129, with G1 located approximately 120\arcsec$\,$ away from the BCG.}\label{fig:lens}
  \end{figure*}

  The lens galaxies presented in this paper, hereafter referred to as G1 and G2, are spectroscopically confirmed members of one of the CLASH galaxy clusters, i.e. MACS J1115.9$+$0129. G1 is the main lens around which four images of a single background source are visible.The potentials of both G2 and the galaxy cluster as a whole also contribute significantly to the displacement of the multiple images. Figure~\ref{fig:lens} shows the optical observations taken with the Subaru and \emph{HST} telescopes and Table~\ref{tab:positions} gives the coordinates and spectroscopic redshifts of all objects relevant to the system. G1 is approximately 120\arcsec$\,$ (i.e., 600 kpc) away from the cluster centre, assumed here to be coincident with the luminosity centre of the Brightest Cluster Galaxy, BCG. The average distance of the four multiple images from the centre of G1 is $\tilde{R}_{\text{Ein}}=2.47$ kpc and this value is used as an effective Einstein radius in the following, when looking at the different lensing models. The BCG is located at R.A.$_{\text{J2000}}=$ 11:15:51.90 and Dec.$_{\text{J2000}}=$ +01:29:55.0 and the cluster redshift is $z=0.352$.

  \begin{table}
    \centering
    \caption{The coordinates and spectroscopic redshifts of G1, G2 and the multiple images (A, B, C and D). The coordinates of G1, the BCG and the multiple images were estimated using the \emph{HST/ACS} observations, whereas Subaru observations were used for G2.}\label{tab:positions}
    \begin{tabular}{cccccc}
      \hline
      & R.A.$_{\text{J2000}}$ & Decl.$_{\text{J2000}}$ & $x_1$ ($'')^{a}$& $x_2$ ($'')^{a}$& $z_\text{spec}$\\
      \hline
      G1      & 11:15:51.63 & +01:31:55.8 &  0.0\,\,\,\,  &  0.0\,\,\,\, & 0.353\\
      G2      & 11:15:51.70 & +01:31:58.1 & -1.01\,\,\,   &  2.30\,      & 0.362\\
      BCG     & 11:15:51.90 & +01:29:55.0 & -4.22\,\,\,   & -121         & 0.352\\
      A       & 11:15:51.68 & +01:31:55.8 & -0.708        &  \,\,0.008   & 2.387\\
      B       & 11:15:51.65 & +01:31:56.2 & -0.299        &  \,\,0.399   & 2.387\\
      C       & 11:15:51.61 & +01:31:56.3 &  \,\,0.274    &  \,\,0.469   & 2.387\\
      D       & 11:15:51.62 & +01:31:55.6 &  \,\,0.129    & -0.204       & 2.387\\
      \hline
    \end{tabular}
    \flushleft
    \scriptsize{$^{a}$ Distance, relative to G1.}
  \end{table}

  The images used to analyse this system are CLASH \emph{HST/ACS} and ground-based Subaru data. The \emph{HST/ACS} mosaics were all produced using procedures similar to those described in \citet{Koekemoer11}, including additional processing beyond the default calibration pipelines to remove low-level detector signatures, as well as astrometric alignment across all filters to a precision of a few milliarcseconds using several hundred sources in each exposure, and final combination of all the exposures into a full-depth mosaic for each filter. Further details about the \emph{HST} and Subaru data are described in \citet{Postman12}, which also includes a general description of the CLASH programme.

  The \emph{HST} observations of MACS~J1115.9+0129 were taken in cycle 19 for a total of 20 orbits. Our strong lensing system is visible in only 7 of the available 16 filters (i.e., f435w, f475w, f606w, f625w, f775w, f814w and f850lp). The pixel size of the \emph{HST/ACS} mosaics is 0.065$''$ and all of the multiple images (A, B, C and D) are resolved, see Figure~\ref{fig:lens}. Since G2 lies outside the \emph{HST/ACS} field of view, we use observations of G2 obtained from the Subaru telescope in the $B$, $V$, $R_{\text{c}}$, $I_{\text{c}}$ and $z$ bands, with a pixel size of $0.2''$ and seeing conditions of approximately $1''$.

\begin{figure}
    \centering
    \includegraphics[width = 0.8\columnwidth]{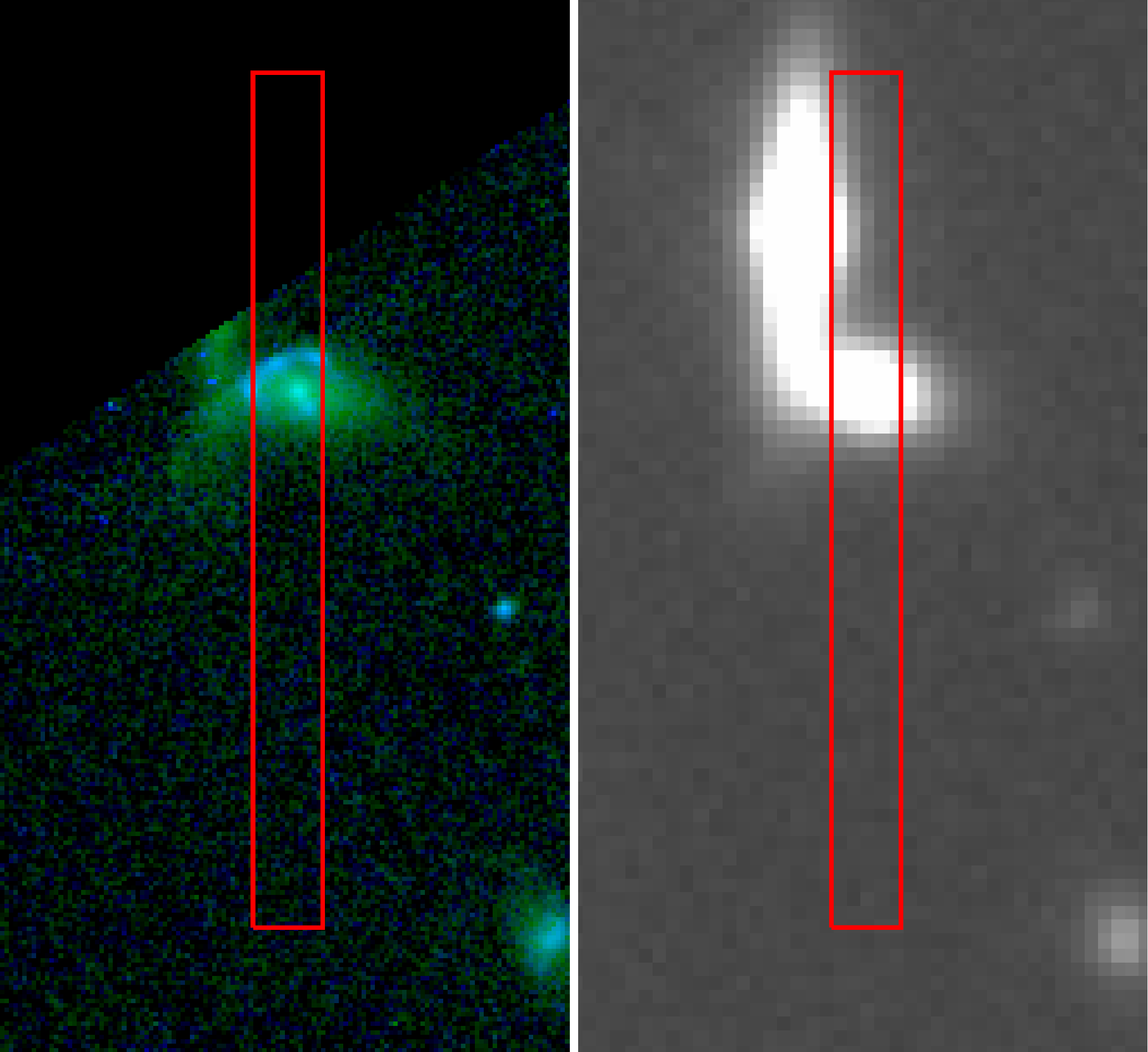}
    \caption{The 1$''$-wide slit position of the CLASH-VLT data obtained with the VIMOS instrument aligned with the \emph{HST/ACS} (\emph{left}) and Subaru (\emph{right}) observations. The extracted spectroscopic data is shown in Figure~\ref{fig:spectra}.}\label{fig:slitalign}
  \end{figure}

\begin{figure*}
    \centering
    \includegraphics[width = 0.8\textwidth]{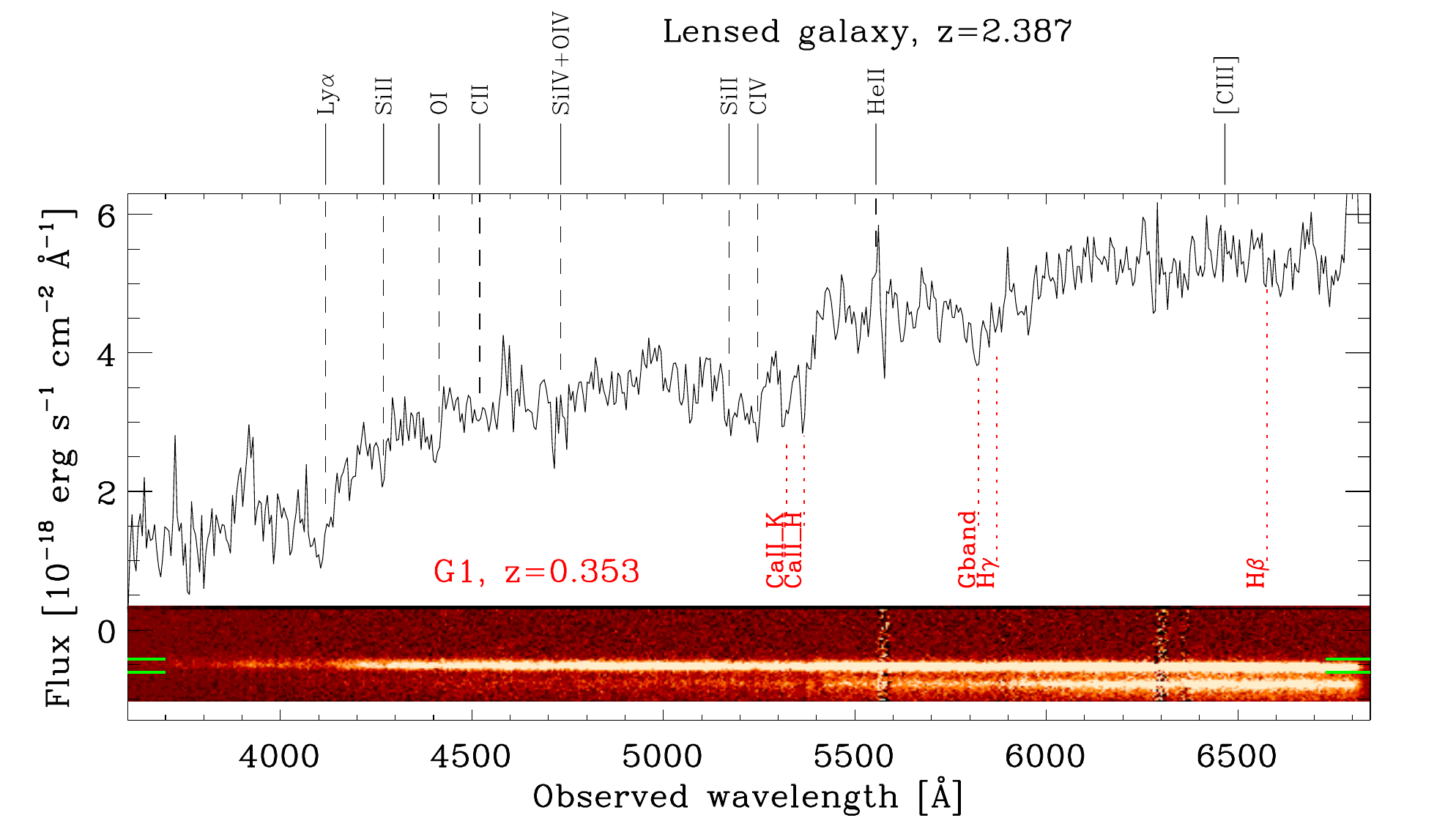}\\
    \includegraphics[width = 0.8\textwidth]{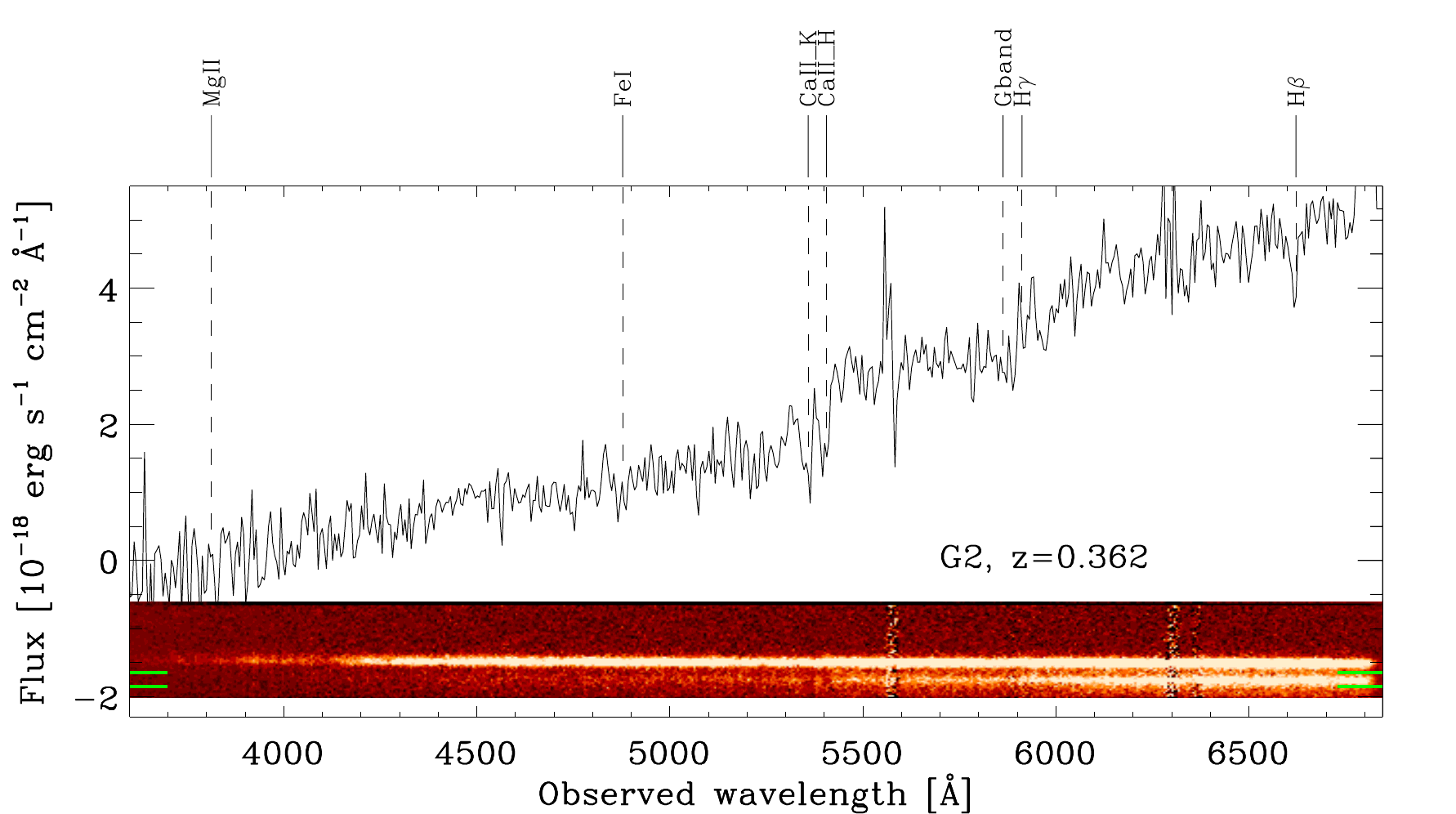}
    \caption{The CLASH-VLT 1-hour spectra, taken with the low resolution blue grism of VIMOS, of G1 and the lensed galaxy (\emph{top}), and G2 (\emph{bottom}). Redshifted features are overlaid on the 1D spectra for G1 (\emph{top, red}) and the lensed galaxy (\emph{top, black}), and G2 (\emph{bottom, black}). The green horizontal lines on the 2D spectra shown beneath each plot locate the position of the 1D spectra.}
    \label{fig:spectra}
  \end{figure*}

  The spectroscopic redshifts presented in this paper were obtained from the CLASH-VLT data \citep{Rosati14} taken with the VIMOS instrument and using the low resolution blue grism, which covers a wavelength range between 3700 and 6700 \AA. The observations have a total exposure time of 2 hours (2 $\times$ 1-hour pointings) and the slit position on G1 is shown in Figure~\ref{fig:slitalign}. In Figure~\ref{fig:spectra}, the extracted 1D and 2D spectra of G1 and G2 and the 1D spectrum of the lensed source are shown, taken from the first 1-hour pointing. In the second pointing the seeing was significantly worse, therefore to maximise the signal-to-noise ratio we preferred to use only the spectra from the first 1-hour pointing. The spectroscopic redshifts of G1 and G2 are 0.353 and 0.362, respectively, and that of the lensed source is 2.387.

\section{Modelling the system}\label{sec:modelling}

  \subsection{Luminosity profiles}

    The luminosity profiles of G1 and G2 are modelled using the publicly available software {\scriptsize GALFIT}, developed by \citet{galfit}. {\scriptsize GALFIT} is an image analysis algorithm which models the light distribution of an object using analytic functions. The values of the adopted model parameters are optimised by means of standard chi-square techniques. To this aim, we created flux error maps from the original drizzle and weight images. The best-fitting parameter values that describe the luminosity profiles of G1 and G2 are shown in Table~\ref{tab:luminosity_profile}. These parameters are the effective radius, $\tilde{R}_{e}$, Sers\'ic index, $n$, major to minor axis ratio, $q_{\ast}$, and the position angle (measured East from North), $\theta_{\ast}$. G2 is modelled with two components: a central bulge and an extended disk, with their centres anchored to each other.

\begin{figure}
      \begin{minipage}{\columnwidth}
        \includegraphics[width = \columnwidth]{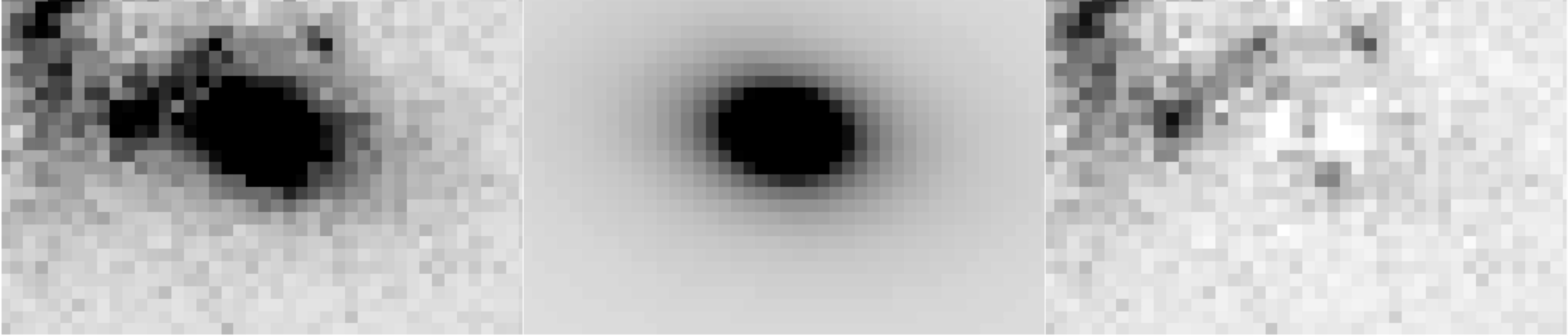}
        \caption{The best-fitting model of the luminosity profile of G1, using \emph{HST/ACS} data in the f850lp band. The original \emph{HST/ACS} image (\emph{left}), the optimised model (\emph{centre}) and the residual after the model subtraction (\emph{right}).}
        \label{fig:galfitG1}
      \end{minipage}
      \begin{minipage}{\columnwidth}
        \includegraphics[width = \columnwidth]{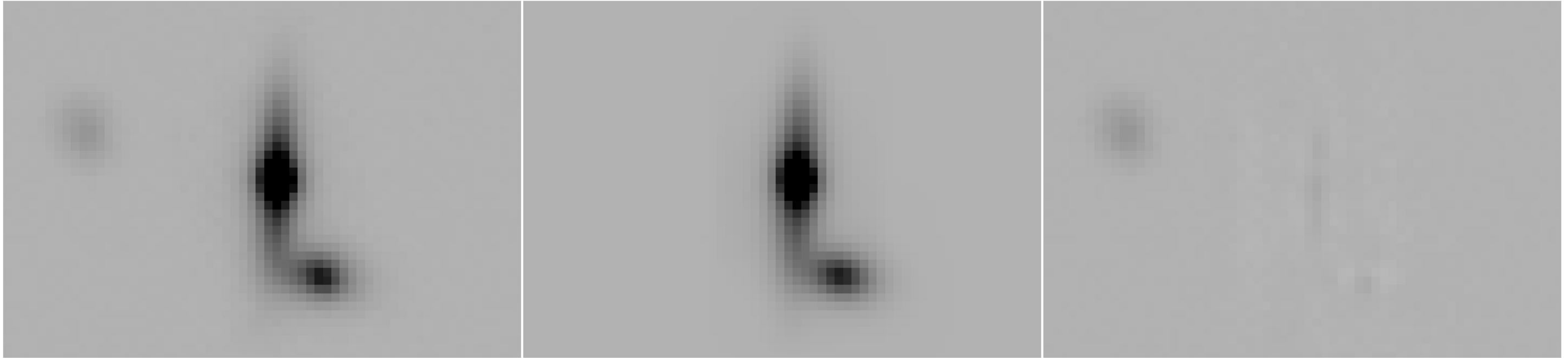}\\
        \includegraphics[width = \columnwidth]{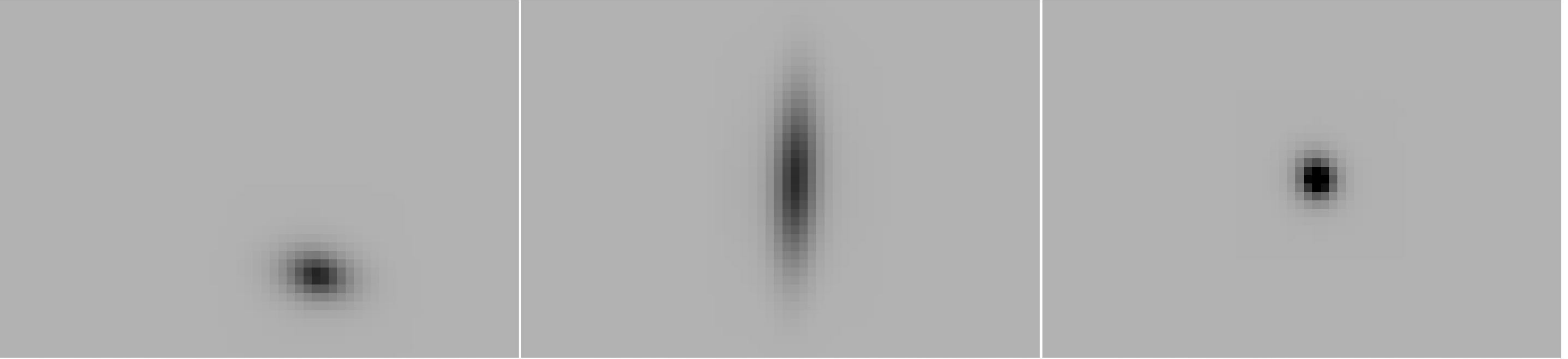}
        \caption{The best-fitting model of the luminosity profile of G2, using Subaru data in the $z$-band. \emph{Top}: original Subaru image (\emph{left}), the optimised model (\emph{centre}) and the residual after the model subtraction (\emph{right}). \emph{Bottom}: the individual components of the model: G1 (\emph{left}), G2 extended disk (\emph{centre}), and G2 bulge (\emph{right}).}
        \label{fig:galfitG2}
      \end{minipage}
    \end{figure}

    For G1, the \emph{HST/ACS} f850lp filter observation is used. To keep the model as simple as possible, the multiple images and the contamination from G2 were masked, and only G1 was modelled. G2 lies outside the \emph{HST/ACS} field of view and therefore Subaru data in the $z$-band is used. We use the reddest bands to model the galaxy luminosity profiles, which minimises the contamination from the multiple images of the bluer, more distant source. A spectroscopically confirmed star is chosen to estimate the PSF for the modelling of G1 and G2. The star coordinates are R.A.$_\text{J2000}=$ 11:15:59.19, Dec.$_\text{J2000}=$ +01:30:22.7, approximately 113\arcsec$\,$ away from the BCG and 147\arcsec$\,$ away from G1. The parameter values shown in Table~\ref{tab:luminosity_profile} are found to be robust when looking in the f775w and f814w images for G1 and the $I_{c}$-band image for G2. The analysis of the ground-based Subaru $z$-band image also provides values of $q_{*,{\rm G1}}$ and $\theta_{*,{\rm G1}}$ that are consistent with those obtained from the \emph{HST} images. In Figures~\ref{fig:galfitG1} and \ref{fig:galfitG2}, we show the reconstructed luminosity profiles of G1 and G2. For both galaxies, we will use the best-fitting values of their respective $q_{\ast}$ and $\theta_{\ast}$ to constrain their total mass profiles in Section~\ref{sec:modellign_lens}. We decide to use the best-fitting values of the extended and prominent disk component for G2.

    \begin{table}
      \begin{minipage}{\columnwidth}
        \centering
        \caption{The best-fitting luminosity profile parameters of G1 and G2 obtained by using {\scriptsize GALFIT}.}\label{tab:luminosity_profile}
        \begin{tabular}{lcccr}
          \hline
          & $\tilde{R}_e$ ($''$) & $n$ & $q_{\ast}$ & $\theta_{\ast}$ ($^\circ$) \\
          \hline
          G1$^{a}$       & $0.86\pm0.04$ & $4.1\pm0.1$ & $0.54\pm0.01$ &  $82\pm1\,\,\,\;$ \\
          G2$_{\text{bulge}}^{b}$ & $0.15\pm0.02$ & $1.2\pm0.9$ & $0.6\pm0.1$ &  $2\pm11\;$ \\
          G2$_{\text{disk}}^{b}$  & $1.60\pm0.02$ & $0.5\pm0.1$ & $0.15\pm0.01$ & $-2\pm1\,\,\,\;$ \\
          \hline
        \end{tabular}
      \flushleft
      \scriptsize{$^{a}$ from the \emph{HST} image in the f850lp filter.}\\
      \scriptsize{$^{b}$ from the Subaru image in the $z$-band filter.}
      \end{minipage}
    \end{table}

    We remark that in the following we will refer to the values of the galaxy half-light radii, $R_{e}$, i.e. the radii inside which half of the total light is contained, which are obtained from the optimised, best-fitting values of Table~\ref{tab:luminosity_profile}. In detail, we find values of $R_{e}$ of 3.3 kpc and 3.1 kpc for G1 and G2, respectively, where the errors on these values are comparable to those of $\tilde{R}_{e}$ and therefore negligible for the lens mass decomposition performed later.

    The luminosity models discussed here for the two lens galaxies will be combined with the Spectral Energy Distribution (SED) fitting results described in the next subsection to estimate the projected luminous over total mass fractions, $f_{\ast}$, presented in Section~\ref{sec:discussion}.

  \subsection{Luminous masses}

\begin{figure}
      \includegraphics[width = \columnwidth]{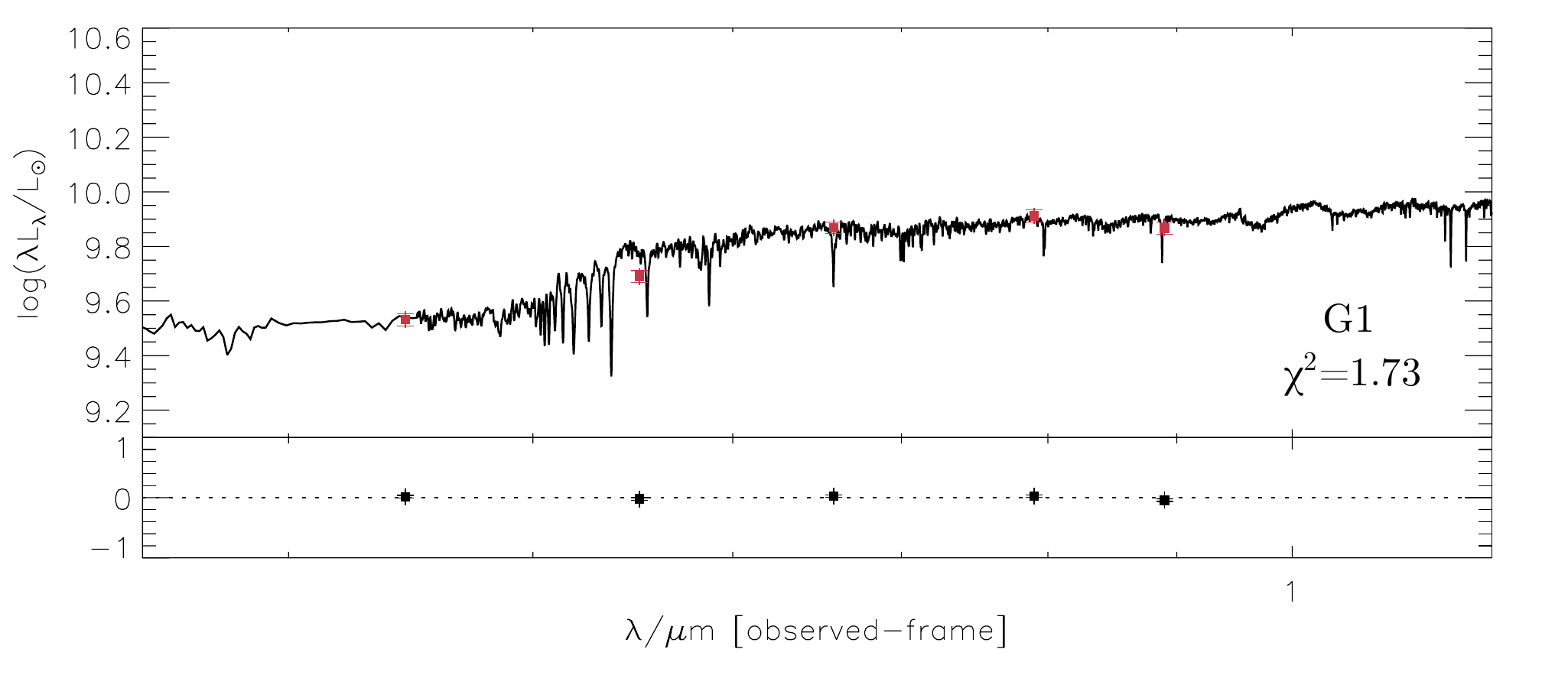}\\
      \includegraphics[width = \columnwidth]{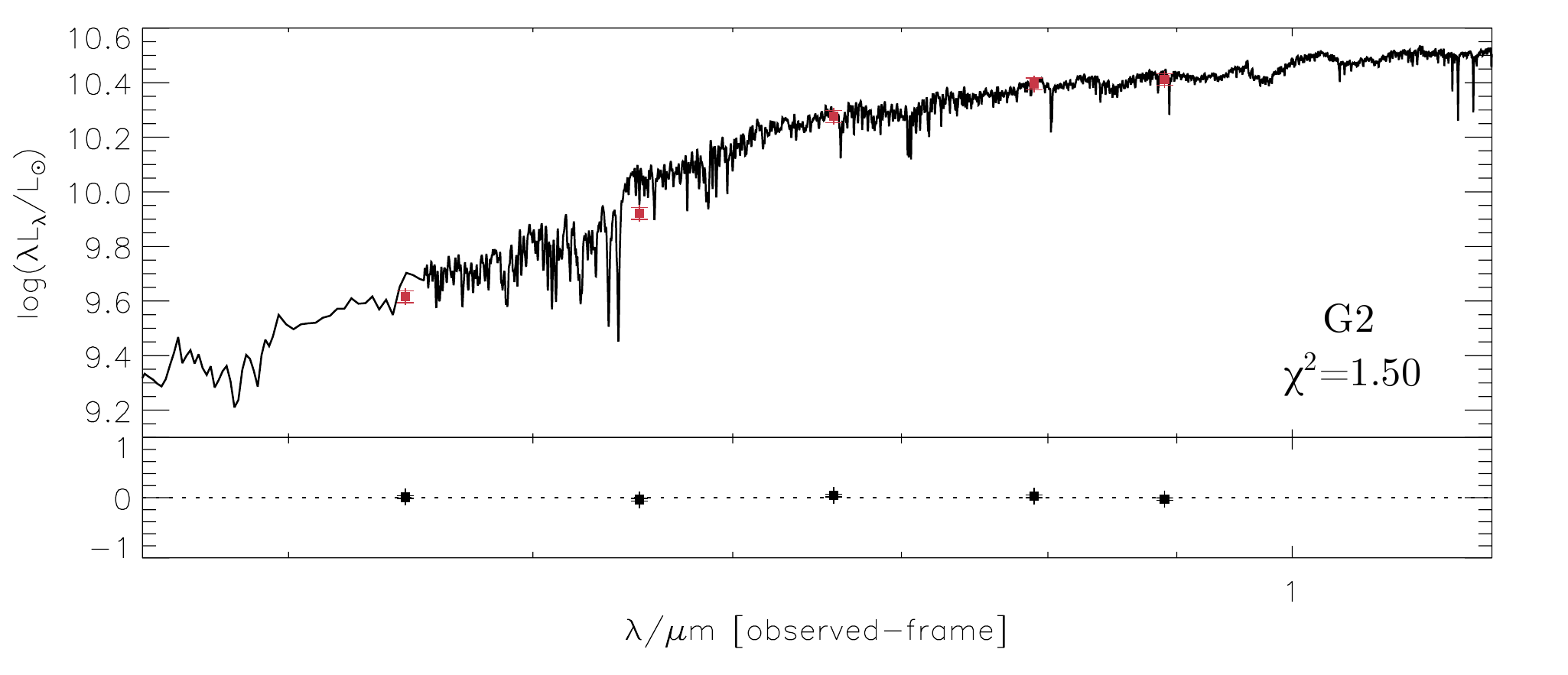}
      \caption{The best-fitting spectral templates used to estimate the total luminous mass values of G1 (\emph{top}) and G2 (\emph{bottom}). For each galaxy, Subaru magnitudes in the $B$, $V$, $R_{c}$, $I_{c}$ and $z$ bands are fitted with stellar population synthesis models, adopting a Chabrier stellar IMF, metallicity values in the range $0.02-2\;Z_{\odot}$ and dust attenuation as prescribed by \citet{CharlotFall00}. The residuals after model subtraction are shown underneath each plot for each wavelength band.}\label{fig:SED}
    \end{figure}

    The photometric magnitudes used to estimate the total luminous mass values of G1 and G2 are from the standard catalogue obtained by the CLASH collaboration and are available in the Subaru $B$, $V$, $R_{c}$, $I_{c}$ and $z$ bands. We model the galaxy SEDs with the {\scriptsize MAGPHYS} code (\citealt{Cunha08}), by using stellar population synthesis models (\citealt{BC03}). We include the effects of dust attenuation, as prescribed by \citet{CharlotFall00}, and adopt a \citet{Chabrier03} stellar Initial Mass Function (IMF) and metallicity values in the range $0.02-2\;Z_{\odot}$. To find the best-fitting model, a Bayesian approach is implemented in {\scriptsize MAGPHYS}. As outputs, the code provides the parameter values of the best-fitting model and the probability distribution functions of each parameter. The best-fitting total luminous mass values of G1 and G2 are ($7.6 \pm 2.3$) $\times10^{9} M_{\odot}$ and ($6.2 \pm 1.2$) $\times10^{10} M_{\odot}$, respectively. The best-fitting template spectra and the relative residuals from the data are shown in Figure~\ref{fig:SED}.

    As shown by \citet{Grillo08c,Grillo09}, the contamination by a blue lensed source does not affect significantly the stellar mass value, estimated through SED fitting, of a red and more luminous lens galaxy. To test this, the value of the magnitude in the bluest band is removed and the SED refitted. We obtain a stellar mass value for G1 that remains within 5\% of the original estimate, where all bands are included. We can therefore conclude that the flux from the lensed images is a secondary source of uncertainty in the measurement of the lens stellar mass, for which errors of approximately 20\% are already considered.

  \subsection{Lens modelling}\label{sec:modellign_lens}

    \begin{table*}
      \centering
      \caption{The results of the bootstrapping analysis with the median values and the 68\% confidence level intervals. The best-fitting parameter values are shown in parentheses.}\label{tab:lensing_models}
      \begin{tabular}{cccccccccc}
        \hline
        Model  & $b_{\text{G1}}$ ($''$) & $q_{\text{G1}}$ & $\theta_{q,\text{G1}}$ ($^\circ$) & $b_{\text{G2}}$ ($''$) & $q_{\text{G2}}$ & $\theta_{q,\text{G2}}$ ($^\circ$) & $b_{\text{C}}$ ($''$)& $\chi^{2}_{\text{tot}}$($N_{\text{dof}}$)\\
        \hline
        SIE\;\;\;\;\;\,\,& \;\;\;\;\;\; - \;\;\;\;\;\;(0.44) & \;\;\;\;\;\; - \;\;\;\;\;\;(0.55)$^{a}$ & \;\;\;\, - \;\;\;\,(35)$^{b}$ & - & - & - & - & 16.9 (3)\\
        \,\,\,SIE+SIS       & $0.32^{+0.04}_{-0.04}$ (0.31) & $0.60^{+0.03}_{-0.04}$ (0.60)$^{a}$ & $88^{+11}_{-9}$ (84)$^{b}$  & $1.14^{+0.13}_{-0.14}$ (1.15) & (1.0) & - & - & 3.58 (2)\\
        \,\;\;\;SIE+2SIS    & $0.32^{+0.04}_{-0.04}$ (0.31) & $0.52^{+0.04}_{-0.04}$ (0.52)$^{a}$ & $76^{+6}_{-4}$ \;\,(75)$^{b}$ & $0.61^{+0.29}_{-0.26}$ (0.66) & (1.0) & - & $31^{+11}_{-12}$ (28) & 0.98 (1)\\
        2SIE+SIS            & $0.32^{+0.04}_{-0.04}$ (0.31) & $0.55^{+0.03}_{-0.03}$ (0.56)$^{a}$ & $76^{+8}_{-7}$ \;\,(75)$^{b}$ & $0.32^{+0.12}_{-0.12}$ (0.32) & (0.2) & (-2.0) & $34^{+8}_{-10}$ (33) & 0.54 (1)\\
        \hline
      \end{tabular}
      \flushleft
      \scriptsize{$^{a}$ Parameter prior of 0.55$\pm$0.1 applied.}\\
      \scriptsize{$^{b}$ Parameter prior of (80$\pm$15)$^{\circ}$ applied.}
    \end{table*}

    To perform our strong lensing analysis, we use the public code {\scriptsize GRAVLENS} \citep{gravlens,gravlens_models}. The total mass profiles of the individual lenses are described in terms of either a Singular Isothermal Sphere (SIS) which is characterised by a single parameter, the lens strength $b$, or a Singular Isothermal Ellipsoid (SIE) which requires two additional parameters: the axis ratio $q$ and the major axis position angle $\theta_{q}$, measured East from North. An isolated lens galaxy modelled with a spherical mass distribution has $b \approx \tilde{R}_{\text{Ein}}$. We remark that the observed $\tilde{R}_{\text{Ein}}$ of a lensing system is in general affected by the lens environment. Therefore, in a galaxy cluster there is a difference between the values of the intrinsic lens strength $b$ of a lens galaxy and of $\tilde{R}_{\text{Ein}}$. In fact, the mass contribution from a cluster or a close galaxy makes the value of $\tilde{R}_{\text{Ein}}$ larger than that of $b$ of the main lens galaxy. {\scriptsize GRAVLENS} uses the value of $b$ to map the convergence, $\kappa$, of a model and the latter can be associated to the total mass of a lens through
    \begin{align}
      \kappa(<R)=\frac{\Sigma(<R)}{\Sigma_{c}},
    \end{align}
    where $\Sigma(<R)$ is the cumulative surface mass density within the radius $R$ and $\Sigma_{c}$ is the critical surface mass density, defined as
    \begin{align}
      \Sigma_{c} = \frac{c^{2}}{4\pi G}\frac{d_{os}}{d_{ol}d_{ls}},
    \end{align}
    where $d_{os}$, $d_{ol}$ and $d_{ls}$, are the observer-source, observer-lens and lens-source angular diameter distances, respectively. The $b$-$\kappa$ relation used by {\scriptsize GRAVLENS} for the mass profiles considered in this work is
    \begin{align}
      \kappa(\zeta)=\frac{b^{2-\alpha}}{2}\left( s^{2}+\zeta^{2}\right)^{\alpha/2-1},
    \end{align}
    where $\alpha$ is a power-law index which is set to 1 for all isothermal models, $s$ is a central core radius at which the model flattens to avoid the singularity at the centre of a SIS/SIE profile, and $\zeta$ describes the elliptical radius in coordinates aligned with the major axis of the ellipse:
    \begin{align*}
      \zeta(x,y) = \left[\left(\frac{2q^{2}}{q^{2}+1}\right)x^{2}+\left(\frac{2}{q^{2}+1}\right)y^{2}\right]^{1/2}.
    \end{align*}

    The multiple images are modelled as point-like objects. In order to determine how well a model reproduces the observations, a positional chi-square value, $\chi_{\text{pos}}^{2}$, is estimated. This is defined as
    \begin{align}
      \chi^{2}_{\text{pos}}=\sum^{N_{\text{I}}}_{i}\frac{||\boldsymbol{x}^{\text{obs}}_{i}-\boldsymbol{x}_{i}||^{2}}{\sigma_{\boldsymbol{x}_i}^{2}},
    \end{align}
    where $N_{\text{I}}$ is the number of multiple images, and, for the $i$-th image on the image plane, $\boldsymbol{x}^{\text{obs}}_{i}$ is its observed position, $\boldsymbol{x}_{i}$ is its model-predicted position and $\sigma_{\boldsymbol{x}_i}$ is the error on the observed position (here chosen to be the same for all the multiple images and equal to the pixel size of the \emph{HST/ACS} mosaics, i.e., $\sigma_{\boldsymbol{x}}=0.065''$).

    In this analysis, the parameters $q_{\ast,\text{G1}}$ and $\theta_{\ast,\text{G1}}$, obtained from modelling the luminosity profile of G1 (see Table~\ref{tab:luminosity_profile}), are used as priors in the lensing models, to avoid unphysical regions of the parameter space. In detail, we consider the following priors: $q_{\text{G1}}=0.55\pm0.1$ and $\theta_{q,\text{G1}}=(80\pm15)^{\circ}$. The 1$\sigma$ errors are chosen large enough not to over-constrain the lensing models. When limits are placed on the model parameters, an additional penalty, $\chi_{\text{prior}}^{2}$, is imposed on the total chi-square value, $\chi_{\text{tot}}^{2}$, according to
    \begin{align}
      \chi_{\text{tot}}^{2} = \chi^{2}_{\text{pos}} + \chi_{\text{prior}}^{2} = \chi^{2}_{\text{pos}} + \sum^{N_p}_{i}\frac{(p_{i}-\tilde{p_{i}})^{2}}{\sigma_{p_{i}}^{2}},
    \end{align}
    where $N_{p}$ is the number of parameters, $p_i$ is the tried value for the $i$-th parameter, $\tilde{p_{i}}$ is the adopted prior on the value of that parameter, and $\sigma_{p_{i}}$ is the 1$\sigma$ error on the prior of that parameter. The best-fitting model parameters are found through minimisation of the $\chi_{\text{tot}}^{2}$ value relative to the number of degrees of freedom, $N_{\text{dof}}$, as the free parameters of a model are varied. The $\chi_{\text{tot}}^{2}$ values of each model, as well as the best-fitting model parameters, can be found in Table~\ref{tab:lensing_models}. The statistical uncertainties are determined for all models through a bootstrapping analysis. We create $10^4$ data sets by sampling random values from Gaussian distributions, taking the values of the observed multiple image positions and the pixel size of the \emph{HST/ACS} mosaics as the mean and standard deviation values. The 68\% confidence level errors, estimated from this bootstrapping analysis, are shown in Table~\ref{tab:lensing_models} and in Figures~\ref{fig:mass_profiles} and \ref{fig:degen}.

    We consider four different mass models in this paper. The simplest model comprises only of G1, modelled as a SIE. The free parameters of this model are $b_{\text{G1}}$, $q_{\text{G1}}$, $\theta_{q,\text{G1}}$, $y_{1,s}$ and $y_{2,s}$, where $y_{1,s}$ and $y_{2,s}$ denote the position of the source on the source plane. All parameters are optimised, and the resulting $\chi_{\text{tot}}^{2}$ of this model is $\chi_{\text{tot}}^{2}(N_{\text{dof}})=16.9(3)$. Since we know that this model is a first, crude representation of the total mass distribution of the deflector and given the thus expected poor reconstruction of the observables, we decide to show only the optimised values of the model parameters (see Table~\ref{tab:lensing_models}) before proceeding with the inclusion of the secondary lens, G2.

    The second model we consider has two mass components: the main lens, G1, and the secondary lens, G2, centred on the corresponding luminosity centres. Here, G1 is described as a SIE and G2 as a SIS. Therefore, the free parameters of this model are $b_{\text{G1}}$, $q_{\text{G1}}$, $\theta_{q,\text{G1}}$, $b_{\text{G2}}$, $y_{1,s}$ and $y_{2,s}$. All parameters are optimised, leaving the model with $N_{\text{dof}}=2$. The overall $\chi_{\text{tot}}^{2}$ of this model is $\chi_{\text{tot}}^{2}(N_{\text{dof}})=3.58(2)$, resulting in a good reconstruction of the multiple image positions.

    \begin{figure}
      \centering
      \makebox[\columnwidth][c]{
        \centering
        \includegraphics[width = 1.2\columnwidth]{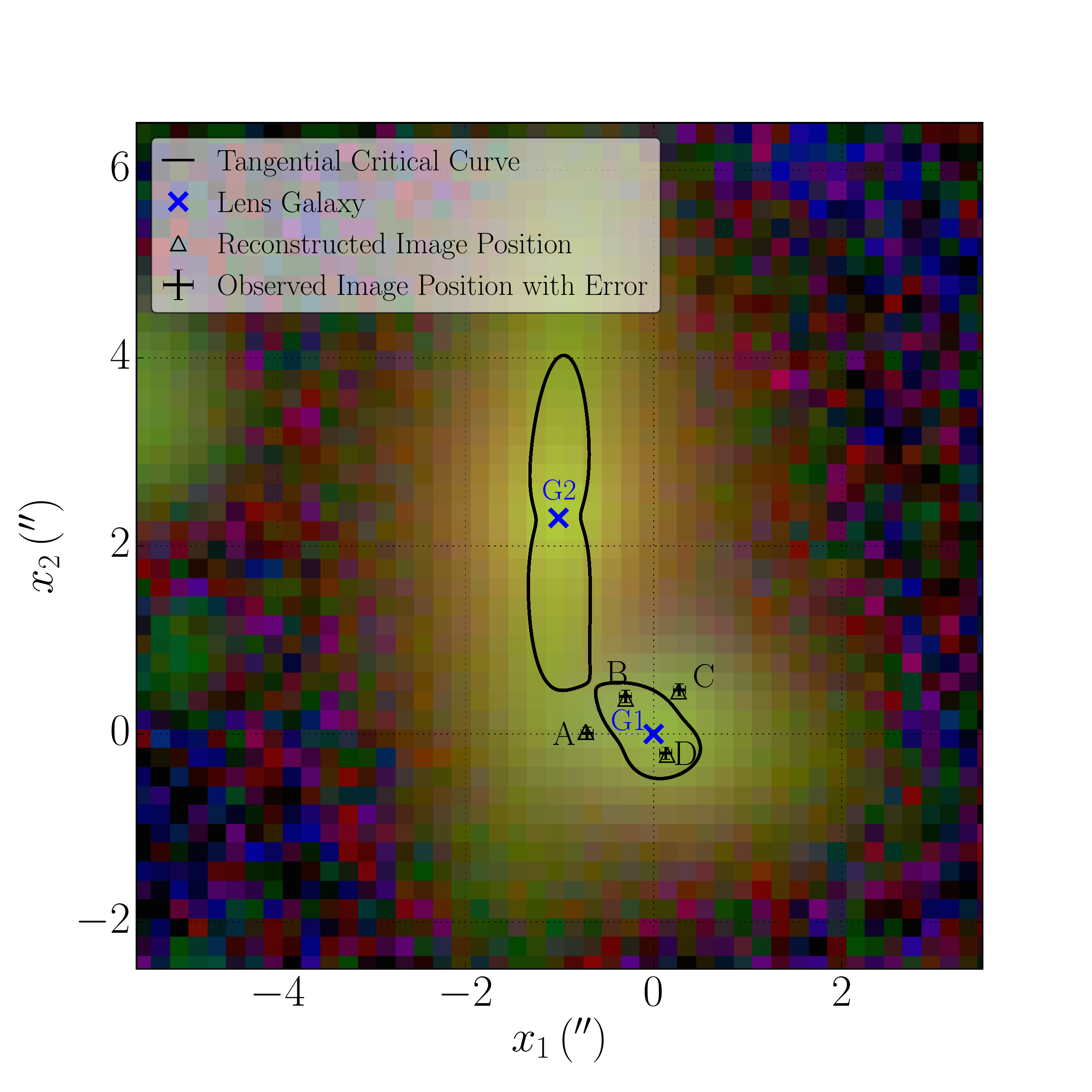}}
      \caption{Multi-band Subaru image with the details of the lensing model obtained for the best-fitting model, where G1 and G2 are described as SIEs and the cluster component as a SIS.}\label{fig:critical_SIE_SIS}
    \end{figure}

    The cumulative projected total mass profile from the centre of G1 and obtained with this model is shown in the top panel of Figure~\ref{fig:mass_profiles}. In this model, the projected total mass value within $\tilde{R}_{\text{Ein}}$ is $M_{\text{T}}(<\tilde{R}_{\text{Ein}})=(3.6\pm0.2)\times10^{10}M_{\odot}$, with G1 accounting for $74\%$ of the projected total mass, i.e. $M_{\text{G1}}(<\tilde{R}_{\text{Ein}}) = (2.7\pm0.3)\times10^{10}M_{\odot}$, and G2 contributing the remaining $26\%$, i.e. $M_{\text{G2}}(<\tilde{R}_{\text{Ein}}) = (0.9\pm0.1)\times10^{10}M_{\odot}$.

    For the third model, we add the galaxy cluster mass component and approximate it with a simple SIS model. This approximation is justified by the large distance between G1 and the BCG, approximately 600 kpc, relative to the average distance between G1 and the multiple images, $\tilde{R}_{\text{Ein}}\approx2.47$~kpc. Although an accurate estimate of the projected mass of a galaxy cluster is not expected from a single galaxy-scale strong lensing system, it is known that some information about the galaxy cluster mass distribution can be inferred (see, e.g., \citealt{Grillo08b,Grillo14}) from systems of this kind.

    To minimise the overall complexity of the models where the cluster is included, G2 is initially described as a SIS and this translates into just one additional free parameter for the lens strength of the cluster, $b_{\text{C}}$. The model reproduces the positions of the multiple images very well, with $\chi_{\text{tot}}^{2}(N_{\text{dof}})=0.98(1)$. The mass profile of this model is shown in the middle panel of Figure~\ref{fig:mass_profiles}. Within $\tilde{R}_{\text{Ein}}$, the projected total mass value is $M_{\text{T}}(<\tilde{R}_{\text{Ein}})=(3.7\pm0.2)\times10^{10}M_{\odot}$, with G1 contributing $73\%$ of the total, i.e. $M_{\text{G1}}(<\tilde{R}_{\text{Ein}}) = (2.7\pm0.3)\times10^{10}M_{\odot}$, and G2 and the cluster contributing $13.5\%$ each, i.e. $M_{\text{G2}}(<\tilde{R}_{\text{Ein}}) = (0.5^{+0.2}_{-0.2})\times10^{10}M_{\odot}$ and $M_{\text{C}}(<\tilde{R}_{\text{Ein}}) = (0.5^{+0.2}_{-0.2})\times10^{10}M_{\odot}$. Although the mass centre of G2 is much closer than that of the cluster to the observed multiple images, our results show that the cluster mass contribution is not negligible.

    \begin{figure}
      \includegraphics[width = \columnwidth]{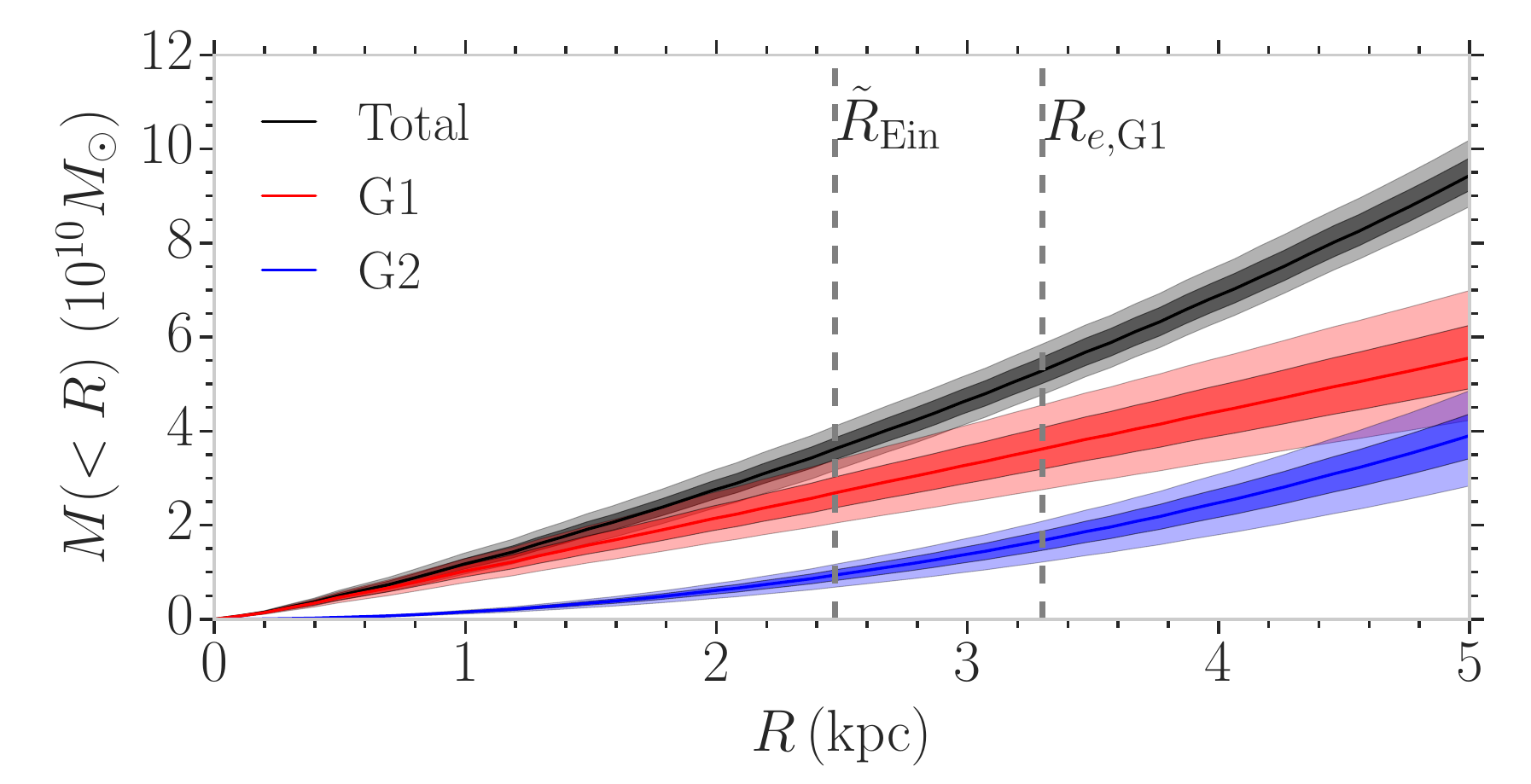}\\
      \includegraphics[width = \columnwidth]{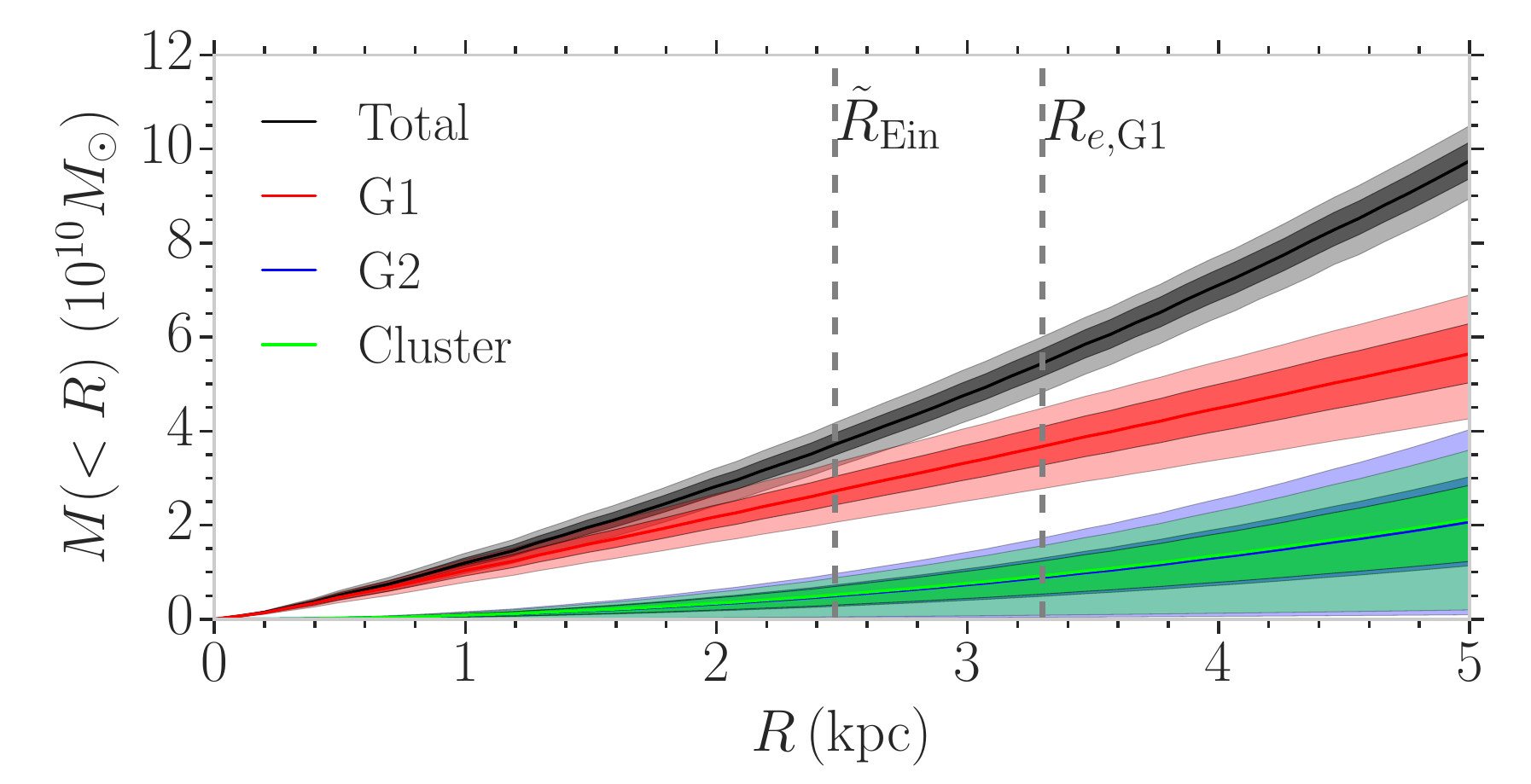}\\
      \includegraphics[width = \columnwidth]{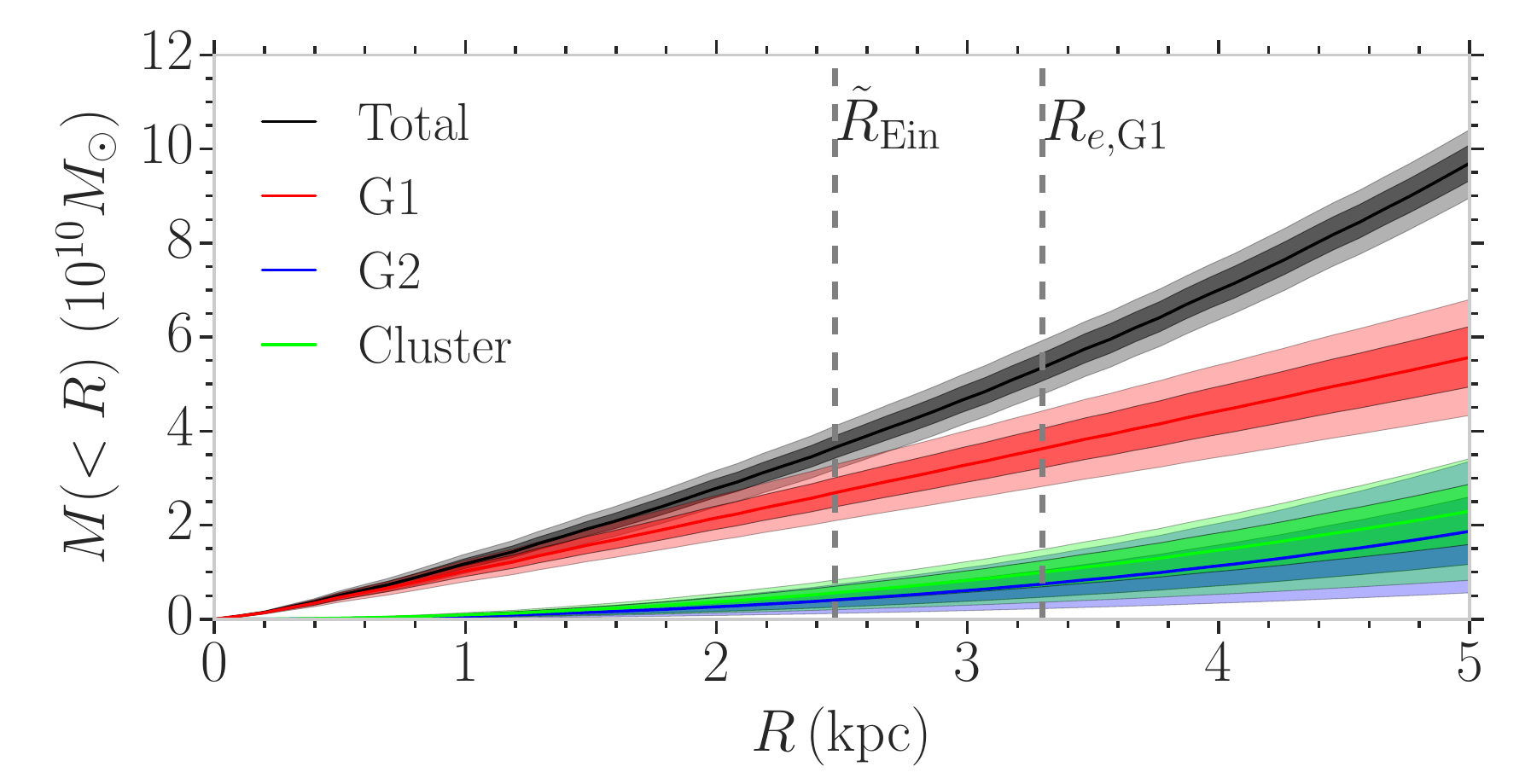}
      \caption{Cumulative projected total mass profiles from the centre of G1 obtained for the different lensing models: SIE+SIS (\emph{top}), SIE+2SIS (\emph{middle}), 2SIE+SIS (\emph{bottom}). The dashed lines show the positions of $\tilde{R}_{\text{Ein}}$ and $R_{e,\text{G1}}$.}
      \label{fig:mass_profiles}
    \end{figure}

    The final model describes both G1 and G2 as SIEs, with the total mass axis ratio and position angle of G2 fixed to its luminosity values, and the cluster as a SIS. This model is found to best reproduce the positions of the multiple images, with $\chi_{\text{tot}}^{2}(N_{\text{dof}})=0.54(1)$. The reconstructed mass profile is shown in the bottom panel of Figure~\ref{fig:mass_profiles}. Within $\tilde{R}_{\text{Ein}}$, the cumulative projected total mass profile is $M_{\text{T}}(<\tilde{R}_{\text{Ein}})=(3.7\pm0.2)\times10^{10}M_{\odot}$. In this model, G1 contributes for $74\%$ of the total mass budget, i.e. $M_{\text{G1}}(<\tilde{R}_{\text{Ein}}) = (2.7\pm0.3)\times10^{10}M_{\odot}$, G2 for $11\%$, i.e. $M_{\text{G2}}(<\tilde{R}_{\text{Ein}}) = (0.4^{+0.2}_{-0.2})\times10^{10}M_{\odot}$, and the cluster for the remaining $15\%$, i.e. $M_{\text{C}}(<\tilde{R}_{\text{Ein}}) = (0.6^{+0.1}_{-0.2})\times10^{10}M_{\odot}$. The details of this model and the positions of the multiple images are shown in Figure~\ref{fig:critical_SIE_SIS}.

\begin{figure}
      \centering
      \includegraphics[width = \columnwidth]{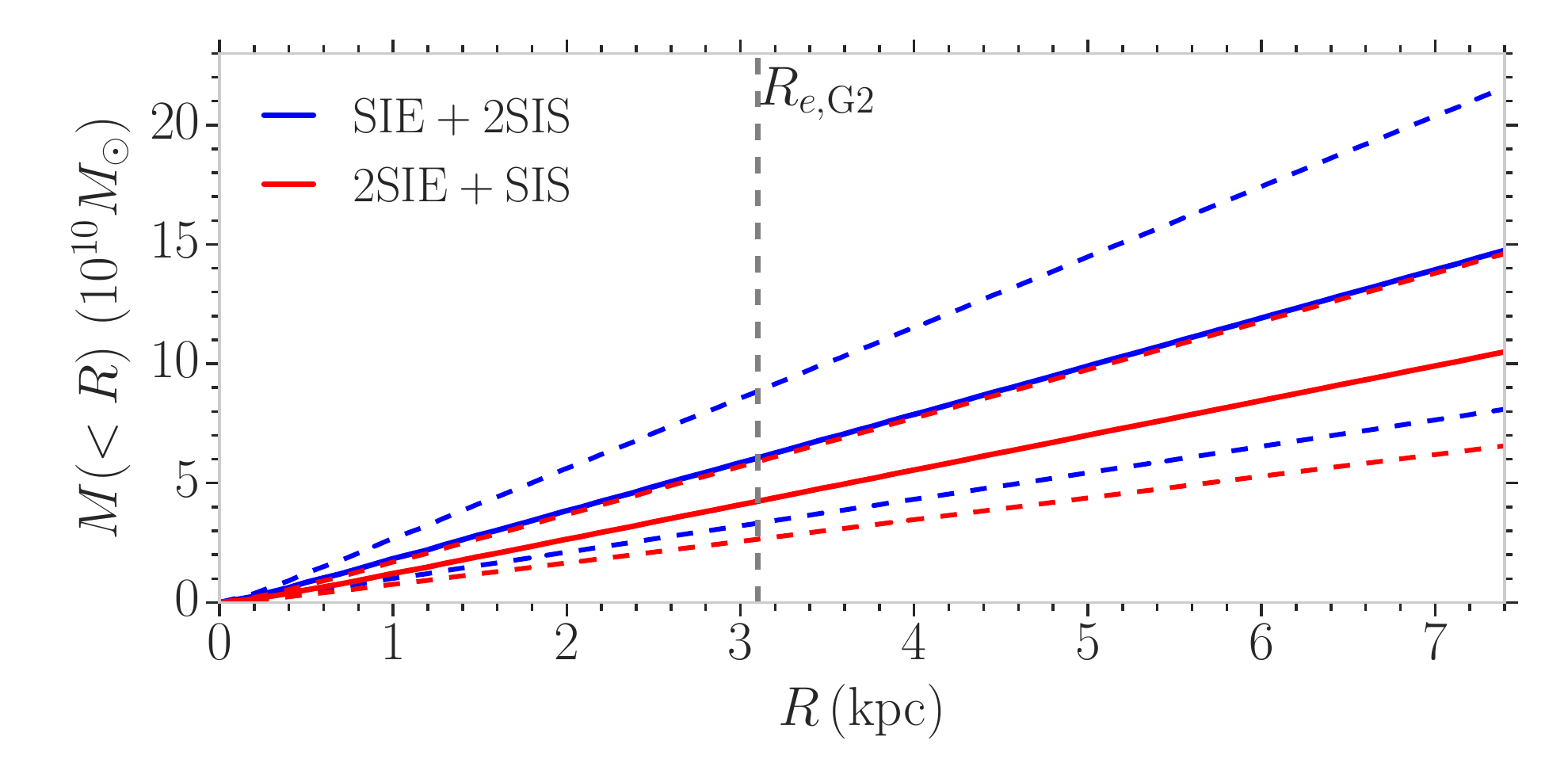}
      \caption{The total mass profile of G2 as measured from its luminosity centre. The solid lines show the median, whereas the dashed lines show the 68\% confidence level intervals obtained from the bootstrapping analysis. The vertical dashed line shows the half-light radius, $R_{e,\text{G2}}$, of G2 estimated from its luminosity profile.}\label{fig:MassProfileAllG2}
    \end{figure}

\begin{figure*}
      \makebox[\columnwidth][c]{
      \includegraphics[scale = 0.45]{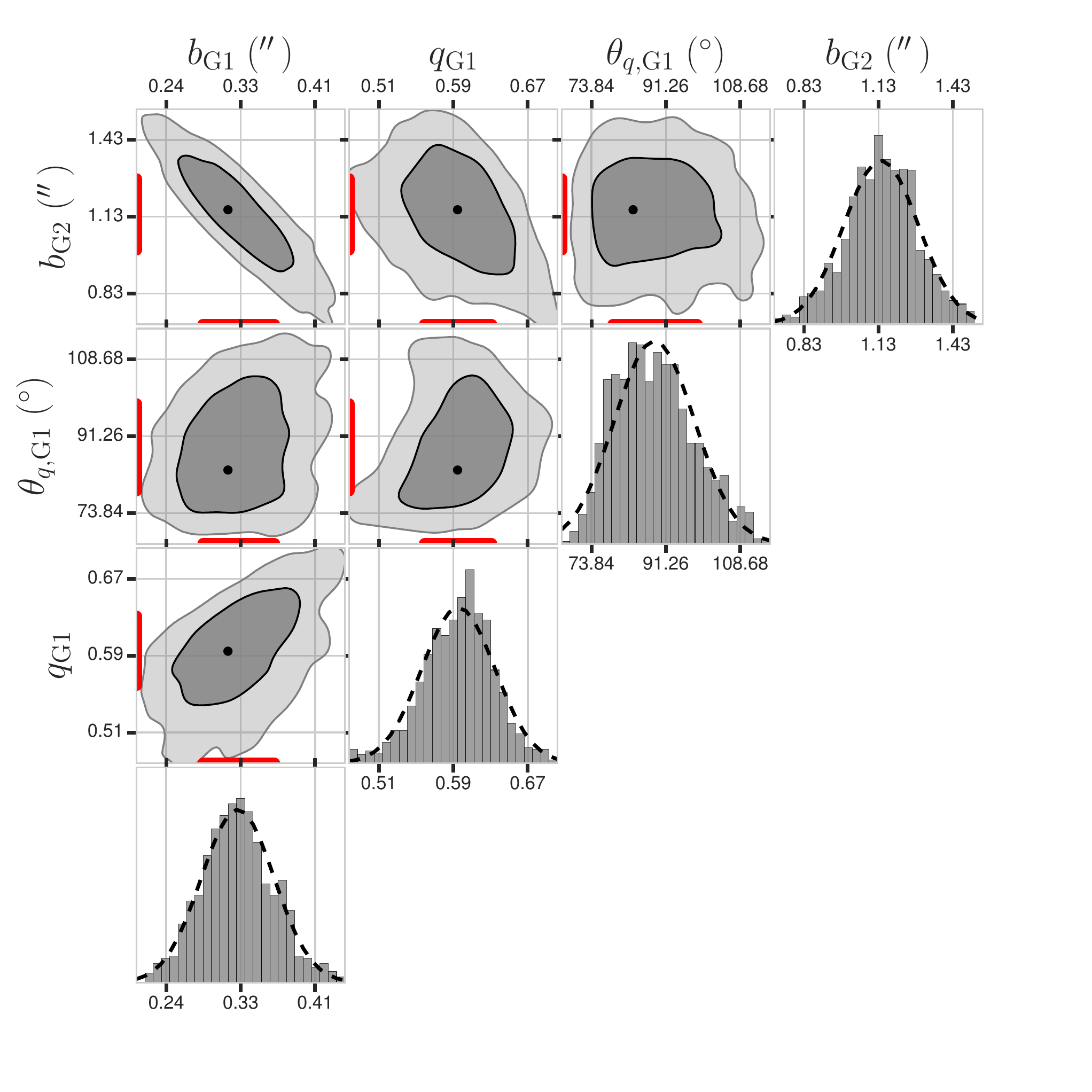}
      }\\
      \makebox[\columnwidth][c]{
      \includegraphics[width = 1.1\columnwidth]{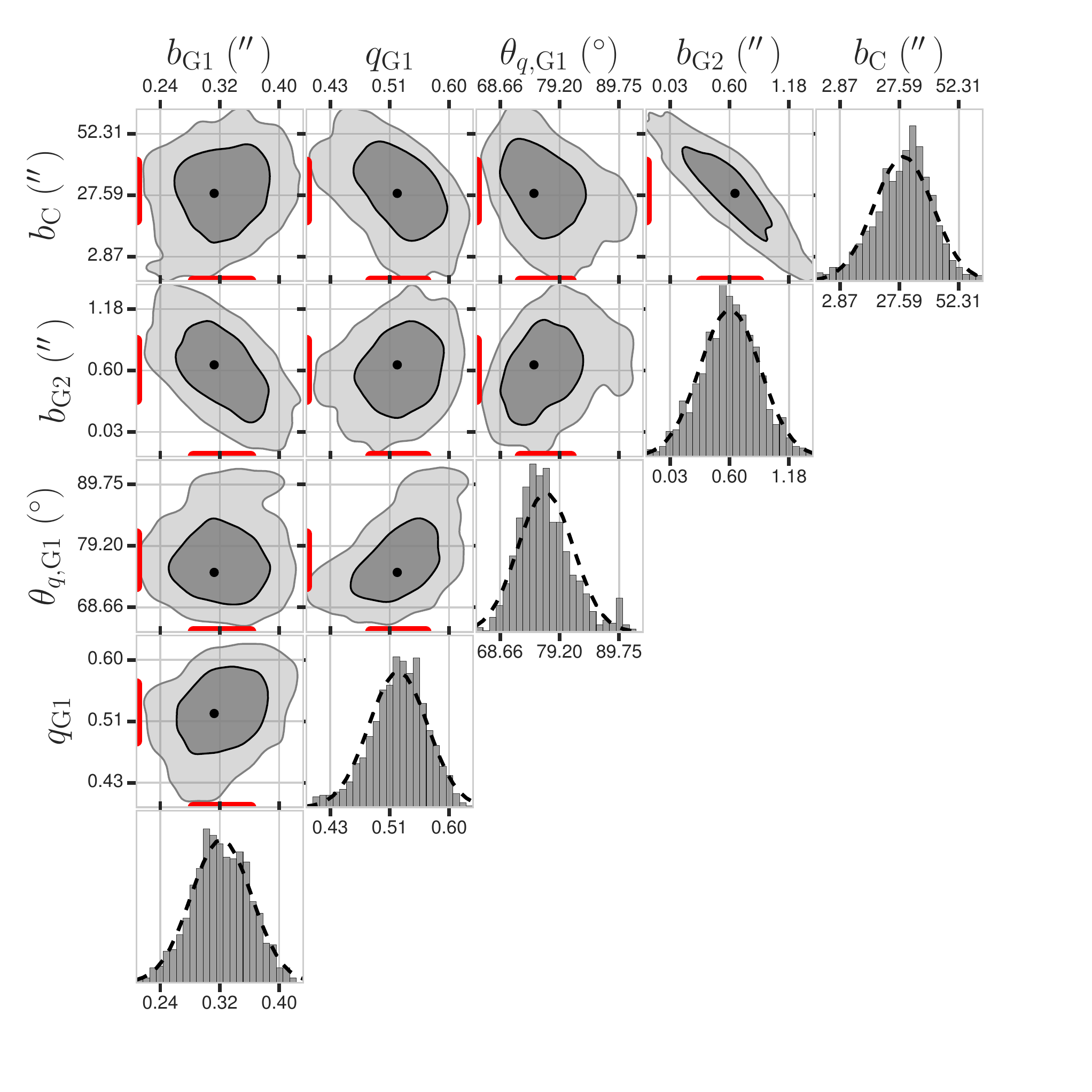}
      \includegraphics[width = 1.1\columnwidth]{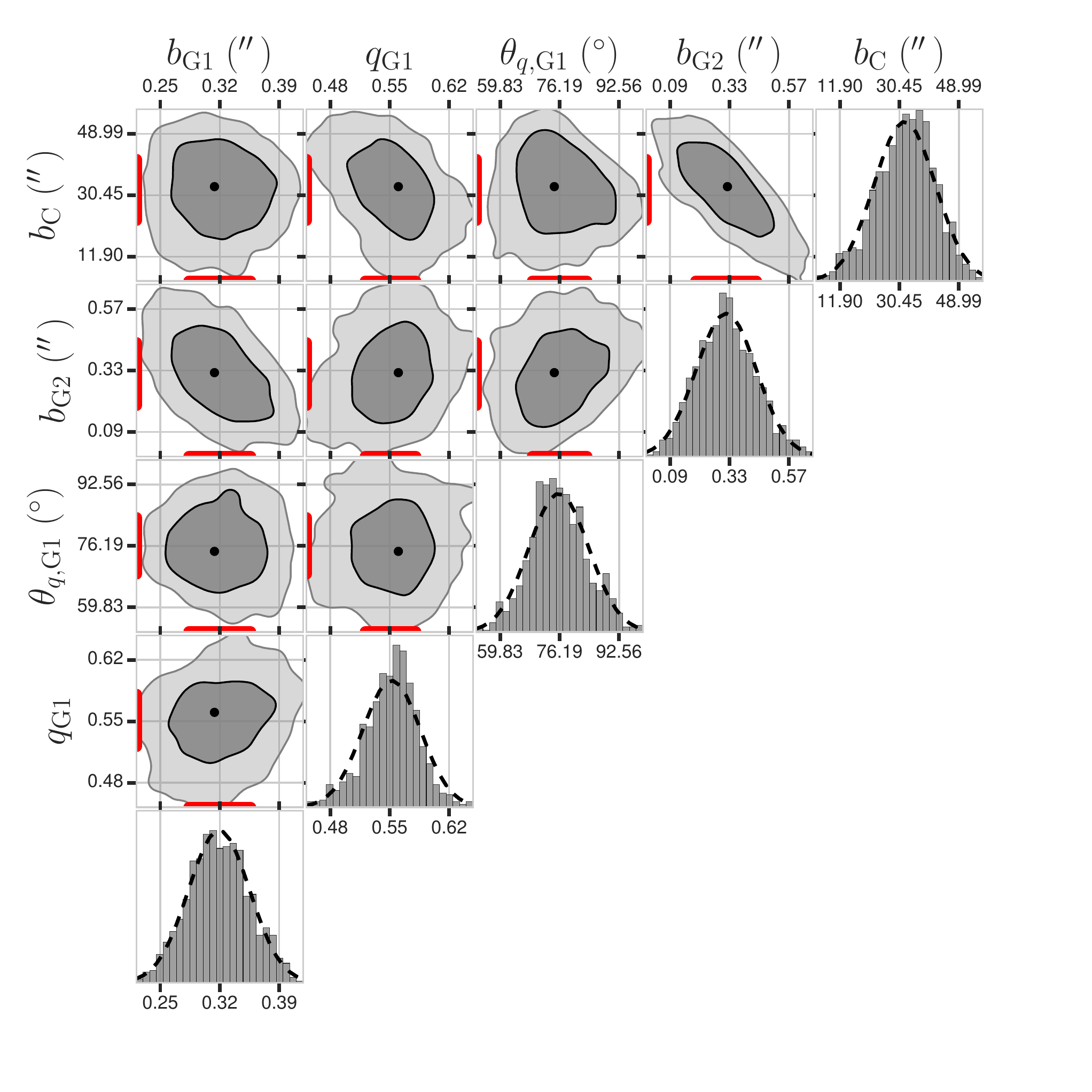}
      }
      \caption{The values of the model parameters obtained from the bootstrapping analysis for the SIS+SIS (\emph{on the top}), SIE+2SIS (\emph{on the bottom left}) and 2SIE+SIS (\emph{on the bottom right}). The black dots show the best-fitting values obtained by optimising over the original multiple image positions. On the axes, the red lines are the 68\% confidence level intervals and the tick marks are the 95\% confidence intervals. On the planes, the contours represent the areas within which 68\% and 95\% of the points are located. The histograms are fitted with simple Gaussian distributions.}
      \label{fig:degen}
    \end{figure*}

    Note that the addition of the cluster mass component does not significantly affect the total mass estimate of G1 (see Figure~\ref{fig:mass_profiles}). On the contrary, the mass contribution of G2 is appreciably lower when the cluster term is present. The lensing observables constrain the projected total mass within $\tilde{R}_{\text{Ein}}$ and the addition of the cluster mass component does not vary this quantity. The total mass is instead redistributed among the three lenses, so that the multiple images are better reproduced. The projected total mass profile of G2, measured from its luminosity centre and derived from the lensing models that include the galaxy cluster component, is also robust, as shown in Figure~\ref{fig:MassProfileAllG2}.

    In general, strong lensing systems found in overdense environments are more complex to model than those associated to a single, isolated lens. As a result, a larger number of parameters is required to reproduce well the observed multiple images. More free parameters usually translate into more degeneracies among them. This is clearly reflected in the system studied here. In Figure~\ref{fig:degen}, we show the degeneracies among the parameters of each model considered in this work. In the first model, where only G1 and G2 are considered, there is an obvious degeneracy between $b_{\text{G1}}$ and $b_{\text{G2}}$. The reason for this is that the total mass found within $\tilde{R}_{\text{Ein}}$ is determined by the total lensing potential. When optimising the model parameters with two mass components the total mass is redistributed between G1 and G2. Once an additional mass component is added this redistribution occurs between all three mass components. In our models, adjusting the mass ratio between G2 and the cluster, through the modelling optimisation, results in a better reconstruction of the multiple image positions, while the mass of G1 remains the same. This is clearly reflected in the values of $b_{\text{G1}}$, $b_{\text{G2}}$ and $b_{\text{C}}$.

\begin{figure}
      \makebox[\columnwidth][c]{
      \includegraphics[width = 1.1\columnwidth]{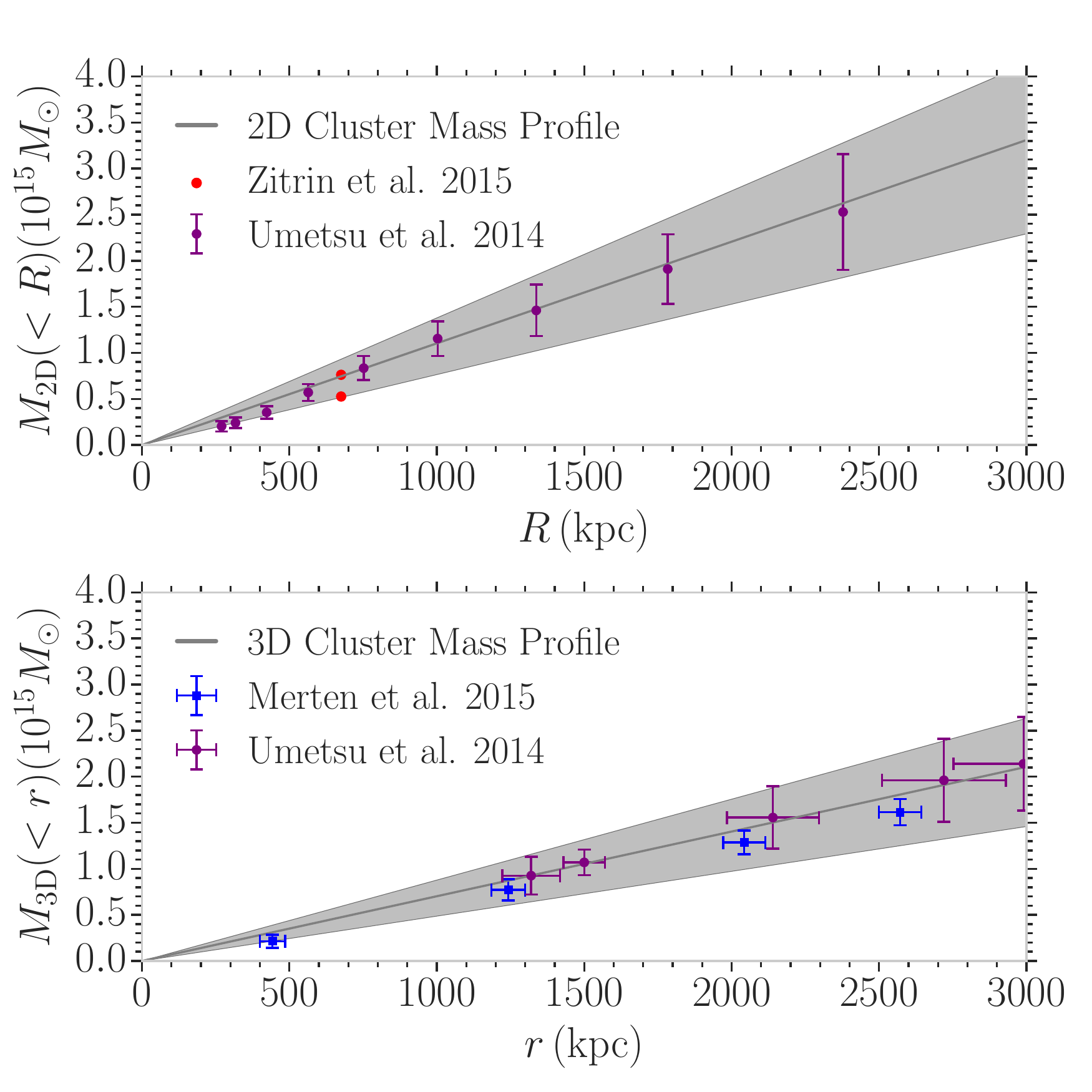}}
      \caption{2D (\emph{top}) and 3D (\emph{bottom}) cumulative total mass distributions of the cluster component from the BCG centre for the best-fitting model, 2SIE+SIS. Here the cluster is modelled as a SIS and the mass distribution is determined purely from the lens strength using equations \ref{eqn:Mass2D} and \ref{eqn:Mass3D}. The errors on the distributions are the 68\% confidence level values obtained from the bootstrapping analysis.}
      \label{fig:MassProfileCluster}
    \end{figure}

    In Figure~\ref{fig:MassProfileCluster}, we show for the galaxy cluster component the reconstructed total 2D mass profile,
    \begin{align}
      M_{\text{2D}}(<R) = \frac{\pi\sigma_{\text{SIS}}^{2}R}{G},\label{eqn:Mass2D}
    \end{align}
    and 3D mass profile,
    \begin{align}
      M_{\text{3D}}(<r) = \frac{2\sigma_{\text{SIS}}^{2}r}{G},\label{eqn:Mass3D}
    \end{align}
    where $\sigma_{\text{SIS}}$ is the value of the effective velocity dispersion of the SIS model, related to the lens strength value $b$ as follows,
    \begin{align}
      \sigma_{\text{SIS}}^{2} = \frac{c^{2}}{4\pi}\frac{d_{os}}{d_{ol}}b.
    \end{align}
    We compare our cluster total mass profiles with those presented in \citet{Zitrin15} and \citet{Merten15}, from a combination of both strong and weak lensing data, and in \citet{Umetsu14}, from a weak lensing study.

    We remark that the 2D and 3D total mass profiles associated to the galaxy cluster component and obtained from our best-fitting model agree very well with the results of the independent analyses mentioned above. This supports the reliability of our strong lensing models, where no information about the cluster total mass was used as a prior.

    \begin{table*}
      \caption{The values of the projected total mass, within the half-light radius, and effective velocity dispersion of G1 and G2.}\label{tab:mass_and_velocity}
      \begin{tabular}{ccccc}
      \hline
      Model             & $M_{\text{T,G1}}(<R_{e,\text{G1}})$ & $\sigma_{G1}$    & $M_{\text{T,G2}}(<R_{e,\text{G2}})$ & $\sigma_{G2}$ \\
                      & ($10^{10}M_{\odot}$)                & (km s$^{-1}$)   & ($10^{10}M_{\odot}$)  & (km s$^{-1}$) \\
      \hline
      \,\,\,SIE+SIS     & $3.6^{+0.5}_{-0.4}$                       & $122^{+7}_{-7}$ & $11.3^{+1.3}_{-1.4}$        & $225^{+13}_{-15}$ \\
      \,\;\;\;SIE+2SIS  & $3.6^{+0.4}_{-0.4}$                       & $122^{+7}_{-7}$ & $\,\,\,6.0^{+2.8}_{-2.7}$   & $164^{+34}_{-43}$ \\
      2SIE+SIS          & $3.6^{+0.4}_{-0.4}$                       & $122^{+7}_{-7}$ & $\,\,\,4.2^{+1.6}_{-1.6}$   & $137^{+25}_{-29}$ \\
      \hline
      \end{tabular}
    \end{table*}

\section{Discussion}\label{sec:discussion}

  In this section, we concentrate on the lens projected luminous over total mass fractions, defined as
    \begin{align}
      f_{\ast}(<R)=\frac{M_{\ast}(<R)}{M_{\text{T}}(<R)}.
    \end{align}

\begin{figure}
      \includegraphics[width = \columnwidth]{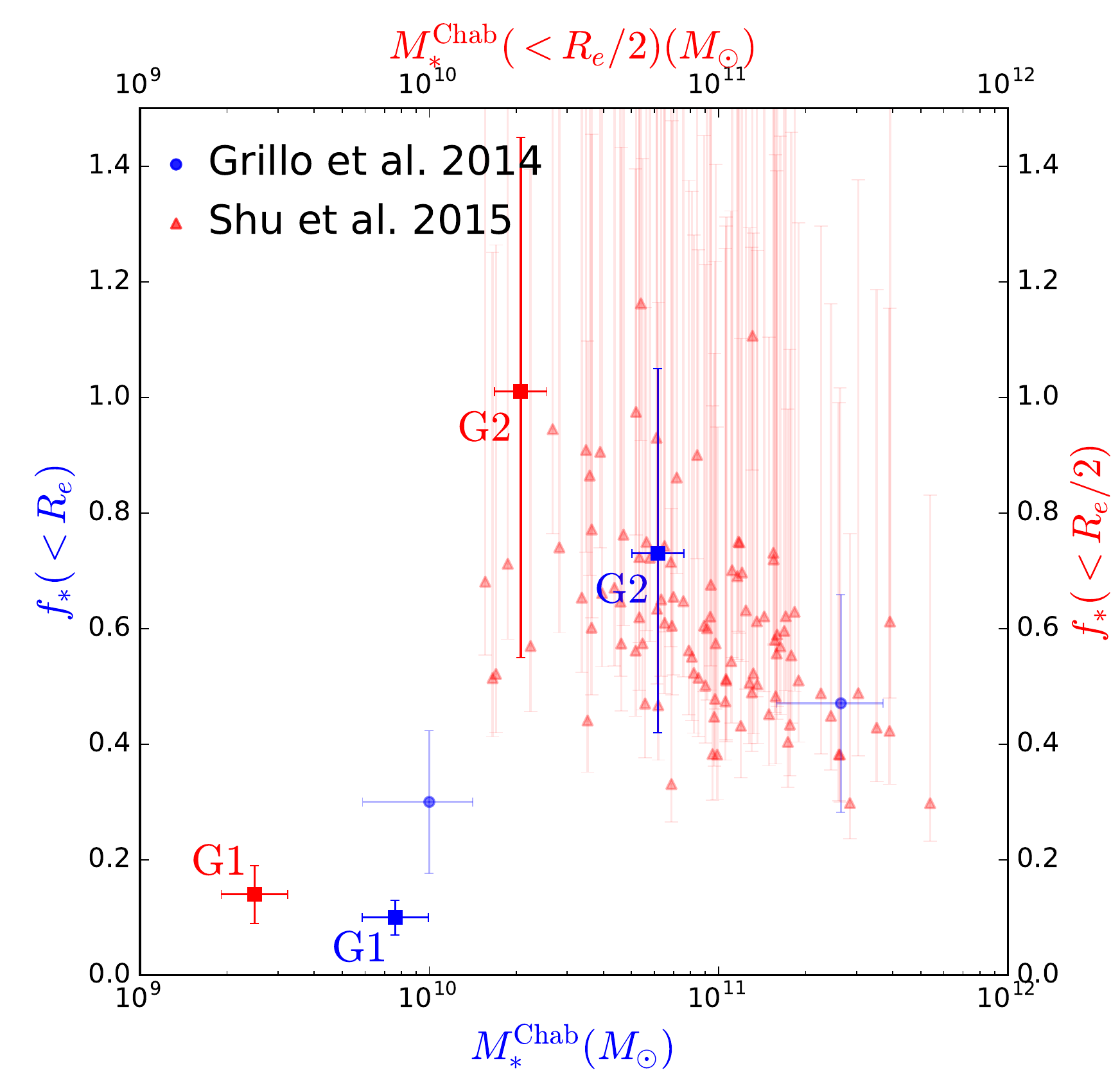}
      \caption{The luminous over total mass fractions measured within $R_{e}$ (\emph{blue}) and $R_{e}/2$ (\emph{red}) plotted against the total luminous mass (\emph{blue}) and the luminous mass within $R_{e}/2$ (\emph{red}). The luminous mass values are shown for a Chabrier stellar IMF. The triangles are the class-A lenses of S15 and the circles are the two lens galaxies investigated by G14. The squares, labelled G1 and G2, are the two lenses studied in this work.}\label{fig:mass_fractions}
    \end{figure}

  The study by \citet{Grillo10}, which considered approximately $1.7\times10^{5}$ massive early-type galaxies from the Sloan Digital Sky Survey Data Release Seven, concludes that the mean value of $f_{\ast}(<R_{e})$ is $0.36\pm0.09$, when a Chabrier stellar IMF is adopted to estimate the galaxy luminous masses. With the same stellar IMF, a more recent analysis by S15 in galaxies with stellar velocity dispersions down to about 140~km~s$^{-1}$ finds that the mean value of $f_{\ast}(<R_{e}/2)$ is $0.60\pm0.16$ (where the quoted error is the standard error of the mean for the 98 class-A lens galaxies in the studied sample, and a class-A lens is defined as a strong gravitational lens with clear and definite multiply lensed images or a complete Einstein ring). Little is known on the values of $f_{\ast}(<R_{e})$ and $f_{\ast}(<R_{e}/2)$ at lower mass scales. One lens galaxy with an effective velocity dispersion of approximately 100 km s$^{-1}$ has been investigated in G14. In Figure~\ref{fig:mass_fractions}, we show a comparison of our $f_{\ast}(<R_{e})$ and $f_{\ast}(<R_{e}/2)$ values with those presented in G14 and S15. It should be noted that in G14 a Salpeter stellar IMF was used. Thus, in Figure~\ref{fig:mass_fractions} the values of that study have been divided by a factor of 1.7 to be converted into the corresponding values for a Chabrier stellar IMF.

  Starting from the total luminous mass values and luminosity profiles obtained in Section~\ref{sec:modelling} and the total mass values listed in Table~\ref{tab:mass_and_velocity}, we estimate the values of $f_{\ast}$ of G1 and G2 within different radii. When comparing with \citet{Grillo10}, we get values of $f_{\ast,\text{G1}}(<R_{e,\text{G1}})=0.11\pm0.03$ and $f_{\ast,\text{G2}}(<R_{e,\text{G2}})=0.73\pm0.32$. When comparing with S15, we obtain values of $f_{\ast,\text{G1}}(<R_{e,\text{G1}}/2)=0.14\pm0.05$ and $f_{\ast,\text{G2}}(<R_{e,\text{G2}}/2)=1.01\pm0.46$. We notice that the projected luminous over total mass fractions of G2 are consistent with those of the galaxies in \citet{Grillo10} and S15. Interestingly, in G1 we find very low values of $f_{\ast}$, suggesting that this galaxy might contain a large amount of dark matter already within its core and significantly deviate from the results obtained for galaxies with masses exceeding $10^{10}M_{\odot}$.

  In Figure~\ref{fig:mass_fractions}, there is a hint of a complex relation between $f_{\ast}$ and $M_{\ast}$. In G14, the comparison of the values of $f_{\ast}$ for the two studied lens galaxies with those of SDSS galaxies and dwarf spheroidals (see the right panel of Figure 10 in that paper) seems to point to a consistent picture. Despite that, with only a few lens galaxies modelled in detail, it is too early to draw any conclusion about the validity of this non-linear relation.

  We notice that cosmological hydrodynamical simulations do not have yet the spatial and mass resolutions needed to estimate reliably the projected luminous over total mass fraction within the half-light radii of cluster members comparable in size to G1 and G2 and residing in galaxy clusters as massive as MACS J1115.9+0129. Interestingly, we remark that, on virial scales, the stellar-to-halo-mass relation reported in abundance matching studies (e.g., \citealt{Moster13,Behroozi13}) seems to show a similar dependence on stellar mass, with a maximum at approximately $10^{10}M_{\odot}$.

  Disparate studies (e.g., \citealt{Treu10b,Spiniello11,Sonnenfeld12,Barnabe13,Newman13b}) agree on finding that a Salpeter-like stellar IMF is the most suitable one for massive early-type galaxies. In the past few years, some observational evidence has also been collected in favour of a non-universal IMF (e.g., \citealt{Thomas11,Spiniello14,Spiniello15}). In particular, the work by \citet{Spiniello15} supports a scenario with a non-universality of the low-mass end of the IMF slope, which decreases with decreasing galaxy mass. Clearly, in our analysis, it is fundamental to know whether a universal stellar IMF is a reasonable assumption \citep{Kroupa01,Cappellari12} or whether we need to adopt a stellar IMF which varies with the galaxy mass.

  Regardless of the choice of the stellar IMF to estimate the luminous mass of the galaxies studied in this paper, our results suggest that the total mass budget in the inner regions of G1 is dominated by the dark matter component. Whether or not this is a characteristic of low-mass galaxies cannot be definitively concluded until more studies on lens galaxies at the same low-mass scale are performed. Strong lensing systems found in overdense environments are inherently more complicated to investigate, however our study shows that with multi-band imaging and spectroscopic data systematics can be controlled and galaxy-scale models are robust. These models can also be further constrained by independent measurements of the cluster total mass component. The study of such systems remains of great importance to extend our knowledge about the internal structure of galaxies at low-mass scales.

\section{Conclusions}

  This paper has presented the strong lensing analysis of a relatively rare system in which the main lens (G1) is a low-mass, cluster member galaxy at $z = 0.353$ which produces four images of a background source at $z = 2.387$. We have shown that an accurate lensing study requires to take into account the mass contributions of a companion cluster member (G2) and of the galaxy cluster (C). In order to extend our knowledge about the internal structure of galaxies to the low-mass end, one possible way is to investigate more complex systems of this kind.

  The main results of this paper can be summarised in the following points:

    \begin{itemize}
      \item The model that best describes the strong lensing system is composed of two SIEs (for the cluster members) and a SIS (for the galaxy cluster), reproducing the observed positions of the multiple images within approximately 0.065\arcsec.
      \item The total mass of G1 projected within its half-light radius is $M_{\text{T,G1}}(<R_{e})=(3.6\pm0.4)\times10^{10}M_{\odot}$, independently of the modelling details.
      \item The total mass of G2 projected within its half-light radius is $M_{\text{T,G2}}(<R_{e})=(4.2\pm1.6)\times10^{10}M_{\odot}$ from the best-fitting model. This quantity shows some degeneracy with the total mass assigned to the galaxy cluster component.
      \item By fitting the SEDs of G1 and G2 with composite stellar population models and assuming a Chabrier stellar IMF, we have obtained that the stellar mass values of G1 and G2 are, respectively, $M_{\ast,\text{G1}}(<R_{e})=3.8^{+1.1}_{-0.9}\times10^{9}M_{\odot}$ and $M_{\ast,\text{G2}}(<R_{e})=3.1^{+0.7}_{-0.6}\times10^{10}M_{\odot}$.
      \item By combining the results of our lensing and photometric analyses, we have estimated that the projected luminous over total mass fractions of G1 and G2 are $f_{\ast,\text{G1}}(<R_{e})=0.11^{+0.03}_{-0.03}$ and $f_{\ast,\text{G2}}(<R_{e})=0.73^{+0.32}_{-0.31}$, respectively.
    \end{itemize}

  Previous strong lensing works have mainly focussed on high-mass lens galaxies in the field, due to the higher probability of being observed and to the specific selection criteria of the past surveys. More recently, some examples of low-mass lens galaxies ($<10^{10}M_{\odot}$), typically in overdense environments, have been detected and studied. The added complexity introduced in these environments requires detailed strong lensing analyses (though limited by the small number of multiple images to only simple mass models), from which relevant information on all contributing mass components can be obtained. The combination of the available results on galaxy-scale strong lensing systems seems to show a variation in the inner luminous over total mass fraction with galaxy mass. The modelling of a larger sample of low-mass lens galaxies, thanks also to exquisite \emph{HST} data collected within the CLASH and Hubble Frontier Fields surveys, is a necessary step to help proving the robustness of these first results and, ultimately, clarifying the precise role played by dark matter in the galaxy mass assembly.

\section*{Acknowledgements}
The authors acknowledge support by VILLUM FONDEN Young Investigator Programme through grant no. 10123. L.~Christensen is supported by DFF - 4090-00079.

\bsp

\label{lastpage}

\end{document}